%% file: main.tex
\documentclass{article} 
\usepackage{iclr2027_conference,times}

\input{math_commands.tex}

\usepackage{hyperref}
\usepackage{url}

\title{\texttt{DivAlgo}: A systematic evaluation of algorithmic diversity for code generation models}

\author{Anubhav Jangra\\ 
Department of Computer Science\\
Columbia University, USA \\
\texttt{anubhav@cs.columbia.edu}
\And
Smaranda Muresan\\
Barnard College \\
Columbia University, USA\\
\texttt{smuresan@barnard.edu}
}

\usepackage{booktabs}
\usepackage{graphicx}
\usepackage{caption}
\usepackage{subcaption}
\usepackage{bbm}
\usepackage{enumitem}

\usepackage[colorinlistoftodos,prependcaption,textsize=tiny,textwidth=14mm]{todonotes}
\makeatletter
\makeatletter
\define@key{todonotes}{AJ}[]{%
    \setkeys{todonotes}{author=AJ,color=cyan!20}}%
\makeatother   
\makeatletter
\define@key{todonotes}{SM}[]{%
    \setkeys{todonotes}{author=SM,color=green!20}}%
\makeatother

\iclrfinalcopy 
\begin{document}

\maketitle

\begin{abstract}
Code generation models are actively being used in professional and educational settings. However, the competence of large language models (LLMs) to generate algorithmically diverse code remains largely unknown. We propose the \texttt{DivAlgo} benchmark, a large-scale dataset constituting algorithmically diverse solutions to programming problems. We also introduce a lightweight similarity metric to enable large-scale evaluation. Our experiments across 24 modern code generation models found that uncontrolled model-generated code is significantly less diverse than human solutions (capturing ${\sim}$31\% of human diversity), and even the best diverse code generation strategy could achieve only ${\sim}$70\% of human diversity. Moreover, while it was possible to steer the models to generate solutions based on human strategies, they were unable to reproduce the diversity of the human solutions. These findings indicate the limitations of code generation models in capturing the algorithmic diversity humans exhibit, and that overexposure to model-generated code can narrow programmers' algorithmic vocabulary.\footnote{Implementation of the work is available at \url{https://github.com/anubhav-jangra/divalgo}.}
\end{abstract}

\input{Sections/1_intro}
\input{Sections/2_related_work}
\input{Sections/4_dataset}
\input{Sections/5_exp_setting}
\input{Sections/6_results}
\input{Sections/8_conclusion}




\subsubsection*{Acknowledgments}
We would like to thank Anshul Nasery for his feedback on the visualizations for this work.

\subsection*{AI use statement}

We have used generative AI tools as aids in writing the code to conduct the experiments, as judges, and in generating the figures and tables. We have not used generative AI for ideation, literature review, or end-to-end paper or section writing. We have reviewed all AI-assisted work, validated the code used for experimentation, cross-checked the illustration artifacts, and ensured that all authors had full agency in the paper-writing process. We take responsibility for the final content of this work, including text, claims, or artifacts produced with the aid of generative AI.

\bibliography{main}
\bibliographystyle{iclr2027_conference}


\appendix
\input{Sections/appx_a_data}
\input{Sections/appx_b_metric}
\input{Sections/appx_c_reproducibility}
\input{Sections/appx_d_results}
\input{Sections/appx_z_prompts}

\end{document}

%% file: math_commands.tex
\usepackage{amsmath,amsfonts,bm}

\def\eqref#1{equation~\ref{#1}}

\def\1{\bm{1}}

\DeclareMathAlphabet{\mathsfit}{\encodingdefault}{\sfdefault}{m}{sl}
\SetMathAlphabet{\mathsfit}{bold}{\encodingdefault}{\sfdefault}{bx}{n}



%% file: Sections/1_intro.tex
\section{Introduction} \label{sec:intro}

Using coding agents and LLMs to perform programming tasks has become ubiquitous not only in software development jobs but also in computer science education and research. 
The swift adoption of these systems has outpaced scrutiny of their outputs, with users overrelying on model-generated code~\citep{shah2025evolution, khati2025mapping}. Much effort to improve LLMs for code generation has focused on executable correctness, using post-training techniques such as reinforcement learning with verifiable rewards (RLVR) that optimize for solution correctness ~\citep{lambert2024tulu, shao2024deepseekmath, wang2026code, li2026exploring, zeng2025acecoder}. However, relying solely on functional correctness as the objective in these learning strategies often concentrates generations on repeated implementations, as examined by ~\citet{florian2026beyond} using a plagiarism detection system for code diversity~\citep{prechelt2002finding}.
If a model is unable to generate the breadth of approaches that humans can develop, then its widespread use narrows the `algorithmic vocabulary' encountered and learned from. Moreover, solution homogeneity has direct engineering costs, as diverse implementations are the foundation of N-version fault tolerance ~\citep{almeida2026effectiveness}. 

Thus, investigating the algorithmic diversity of model generated solutions has become an important area of research with contribution in code similarity \textit{metrics}, 
such as, lexical or embedding overlaps \citep{ren2020codebleu, feng2020codebert}, execution behavior similarity \citep{zhang2026rethinking}, code diversity \textit{benchmarks} \citep{zhang2026rethinking, narula2026the} and \textit{approaches} to increase the diversity of model generated algorithmic solutions \citep{lee2025diversely,florian2026beyond}.
\begin{figure}
    \centering
    \includegraphics[width=0.9\linewidth]{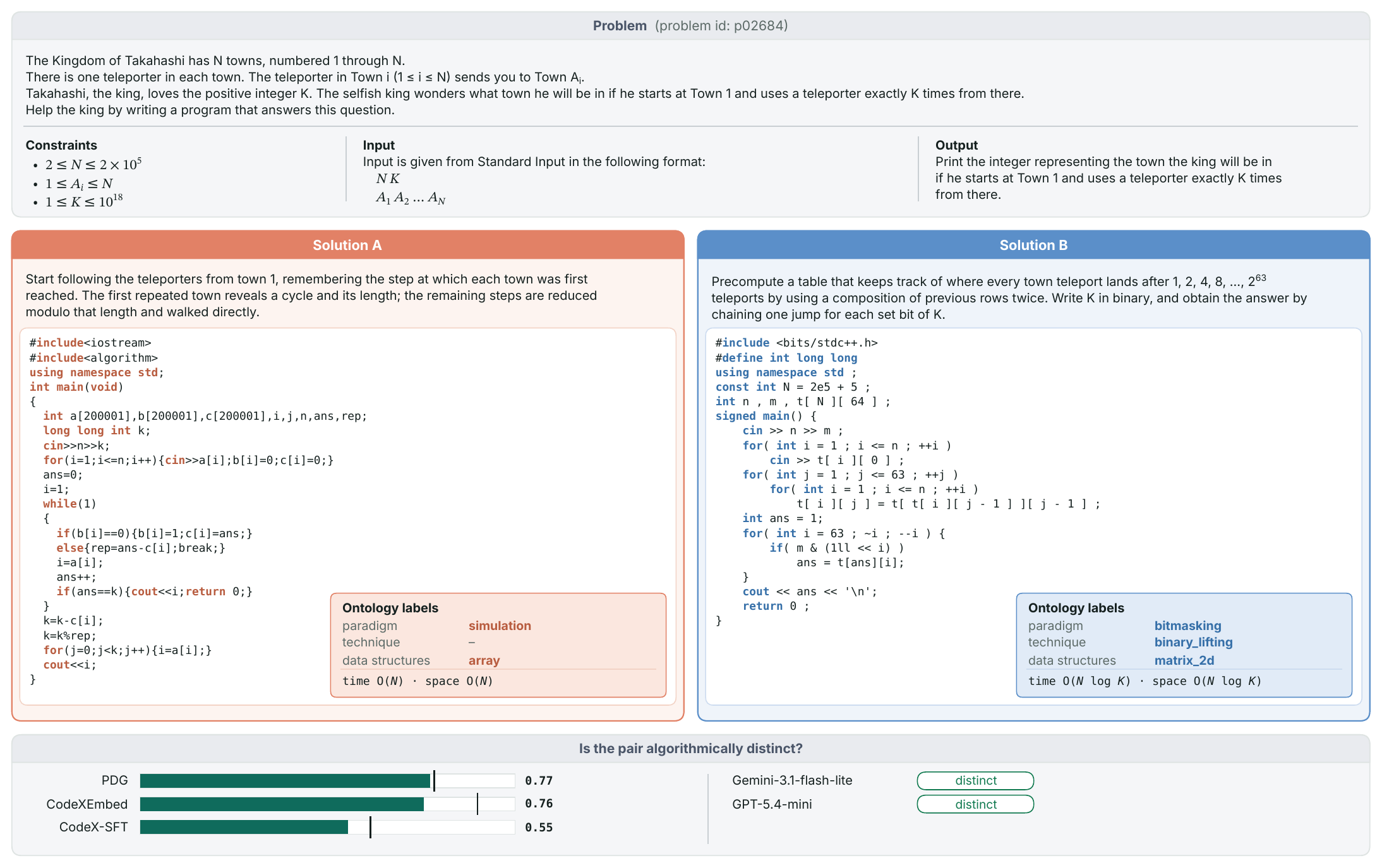}
    \caption{Illustrative example from our algorithmic diversity benchmark's test split demonstrating two algorithmically distinct solutions for the same problem. Solution A solves the problem using the \textit{simulation} paradigm with an \textit{array} data structure, whereas Solution B adopts a \textit{bitmasking} paradigm with the \textit{binary lifting} technique using a \textit{2D matrix} data structure. Both solutions have different time and space complexity. The bar graphs show the code similarity of the two solutions, and the labels show the LLM judgments. The ticks in the bar graph indicate the similarity thresholds, below which the solution pair is considered algorithmically diverse in the respective metric space.}
    \label{fig:example}
\end{figure}
Building on the prior work, we ask three research questions: 
\textbf{RQ1:} How diverse are model-generated solutions compared to human-written solutions? \textbf{RQ2:} How much algorithmic overlap exists between solutions from different models for the same task? \textbf{RQ3:} Can models generate solutions that follow the same strategies adopted by humans? These questions encompass three different notions of algorithmic diversity: within a model, across models, and relative to specific human strategies.

We make three main contributions. \textbf{First}, our new benchmark, \texttt{DivAlgo}, allows us to investigate algorithmic overlap at three levels of granularity: 
data structures, techniques, and programming paradigms. It contains two algorithmically distinct human solutions, along with a pool of human-written solutions for population-level analysis (see Figure~\ref{fig:example} for an example). We systematically constructed the benchmark from \texttt{CodeContest+} \citep{wang-etal-2025-codecontests} via cascaded filtration across distinct code similarity metrics and LLM-judge ensembles, yielding a high-quality, large-scale dataset (Section \ref{sec:dataset}).
\textbf{Second}, to answer the three research questions at scale, 
we introduce a distilled and validated 
lightweight similarity metric and use the Vendi score~\citep{friedman2022vendi} with this metric to estimate the number of distinct algorithms in a population of solutions (Section~\ref{subsec:metric}).  \textbf{Third}, we thoroughly evaluate the capabilities of a large set of LLMs to generate diverse solutions, answering our three research questions. For RQ1, we show that uncontrolled sampling (for 24 LLMs) yields considerably less diverse solutions than human solutions, with the best model achieving 31\% diversity compared to human solutions. Even when using diverse code generation strategies such as temperature variations, verbalized sampling \cite{zhang2025verbalized}, and iterative constrained generation \cite{ye2025assessing}, we can only achieve up to 70.5\% diversity relative to human solutions with an equal number of samples (Section~\ref{sec:rq1}). For RQ2, pooling unconditioned solutions across all 24 models is less effective than using a single model with an appropriate prompting strategy, as we found that the models systematically aligned with the more popular human solution approach (Section~\ref{sec:rq2}). For RQ3, even when the models were steered to generate solutions corresponding to both representative human solutions, they were unable to reproduce the diversity of the overall human solutions (Section~\ref{sec:rq3}).     

%% file: Sections/2_related_work.tex
\section{Related Work} \label{sec:rw}

Code generation has been one of the most popular testbeds for measuring the reasoning capabilities of large language models~\citep{chen2021evaluating}. The verifiable nature of program evaluation makes them ideal for post-training paradigms such as group-relative policy optimization~\citep{shao2024deepseekmath} and  RLVR~\citep{lambert2024tulu}, with prior work developing models capable of program synthesis~\citep{wang2026code, li2026exploring} and automatic test case generation~\citep{zeng2025acecoder}. There exist several benchmarks to evaluate the functional correctness of model-generated code, such as \texttt{HumanEval}~\citep{chen2021evaluating}, \texttt{CodeContests}~\citep{li2022competition}, and \texttt{CodeContests+}~\citep{wang-etal-2025-codecontests}. We refer readers to the extensive surveys for a comprehensive review~\citep{jiang2026survey, dehaerne2022code, chen2024survey}. 

While evaluating the functional correctness of model-generated code remains the most popular focus of prior work, there are some efforts in investigating the diversity of model-generated solutions. ~\citet{lu2025benchmarking} proposed ``\textit{denial prompting}", a systematic constrained generation approach that iteratively limits the techniques models can use to solve the programming problem, eliciting diversity beyond the vanilla model-generated outputs. They evaluate diversity by using the presence of atomic labels (such as \textit{while loop}, \textit{break statement}) obtained with LLM-based classifiers to establish convergent and divergent creativity in the generated solutions relative to previously generated and human solutions, respectively. While they measure LLMs' capabilities in generating diverse solutions, they limit the study to very simple problems (Codeforces 800-level problems) with partially syntax-dependent constraints. \citet{lee2025diversely} adopted a clustering-based approach to identify the number of algorithmically unique solutions using binary classification obtained from \texttt{Llama-3.1-8B}~\citep{grattafiori2024llama}. They found that increasing the temperature and sampling using heterogeneous models can help improve the diversity of model-generated outputs. Complementing them,~\citet{narula2026the} kept the sampling fixed and investigated the impact of varying incidental prompt cues that carry no algorithmic content (such as persona, team names, project codenames, and color themes). They found that these cues can substantially shift the algorithm-family distributions, and the most reliable mitigation strategy was direct algorithm naming. Their approach to detecting algorithm families relied on manually specified abstract syntax tree~\citep{mccarthy1964formal, knuth1968semantics} patterns for each task to reliably capture structural signals, thereby limiting the investigation to 11 tasks. \citet{zhang2026rethinking} instead investigated algorithmic similarity based on execution behavior by executing the programs and capturing the alignment of intermediate states. While precise, this approach cannot be extended to problems without a tractable intermediate structure (see Figure~\ref{fig:example} for an example). Beyond programming problems, \citet{almeida2026effectiveness} connects this research direction with the software-diversity literature, extending classical models of correlated failures~\cite{avizienis2004basic, lyons1962use} to LLM-generated code. They find that heterogeneity across models, temperatures, and programming languages yields measurable reliability gains, but failure independence is not guaranteed with this approach.

Prior work examined the problem of algorithmic diversity for very simple problems~\citep{lu2025benchmarking}, or at a very small scale~\citep{narula2026the}, for a particular class of problems~\citep{zhang2026rethinking}, or at a very coarse level~\citep{lee2025diversely, florian2026beyond}. Our work extends this literature by developing a large-scale algorithmic diversity benchmark spanning problems of varying difficulty, with fine-grained control over the algorithmic overlap between solutions. We distilled a lightweight algorithmic similarity metric to comprehensively examine the diversity of model-generated code relative to human solutions across several code generation strategies.





%% file: Sections/4_dataset.tex
\section{\texttt{DivAlgo} Benchmark Construction} \label{sec:dataset}
We start with the CodeContests+ benchmark, which comprises $11,690$ programming problems from Codeforces, LeetCode, and AtCoder, with $4.36$M correct submissions. To prevent any confounds from inter-language similarity measurements~\citep{utpala-etal-2024-language}, we limit our investigation to the $1.73$M C++ solutions in the dataset\footnote{We decide to use C++ because it is the most popular language in the dataset, spanning $39.7\%$ of solutions, compared to $33.6\%$ for Python-3, $20.6\%$ for Java, and $6.1\%$ for Python-2.}. We filtered out $1,326$ instances with fewer than $20$ correct C++ submissions and dropped $205$ instances without a problem statement. We additionally dropped $966$ duplicate instances\footnote{The Codeforces platform has two divisions, Div1 and Div2, and duplicate problems are used across divisions with different difficulty levels. To only consider unique problems, we dropped duplicate problems by matching normalized prefix hashes over the first 1,000 characters of the problem statement.}. This filtration left us with a canonical set of $9,193$ instances, with a mean of $176.4$ C++ correct submissions. 


To build \texttt{DivAlgo} we need to identify algorithmically different solutions among the human provided solutions. 
A recent human-annotation study \citep{lee2025diversely} has shown the difficulty of obtaining high-quality human annotations for algorithmic diversity, particularly at scale. 
Thus, we conduct a systematic investigation of automatic metrics 
and a two-stage diversity selection process. 
We focus on  
similarity metrics that go beyond 
superficial syntactic attributes (such as variable names): 
1) program dependency graph (\texttt{PDG}) that captures the control and data dependencies for a program, 
2) a code embedding model, 
\texttt{CodeXEmbed-2b} \cite{liu2024codexembed}, 
which was the best model on the code retrieval (CoIR) leaderboard~\cite{li2025coir} in July 2026, and 3) six proprietary LLMs (\texttt{Gemini-3.1-pro}, \texttt{Gemini-3.1-flash-lite}, \texttt{GPT-5.5}, \texttt{GPT-5.4-mini}, \texttt{Claude-opus-4.8} and \texttt{Claude-Haiku-4.5}) used as LLM-as-judge. The models were prompted to give the similarity classification out of four choices: \textit{identical}, \textit{variation},  \textit{distinct} and \textit{very distinct} (prompt in Appendix~\ref{appx:prompts}).

An analysis on 100 randomly selected solution pairs (Appendix~\ref{subsec:preliminary}) reveals that LLM-judges have high pairwise AUC (${\sim}0.85$) despite modest rank correlation, indicating their disagreement arises from the decision boundaries of algorithmic similarity. Additionally, both lightweight measures (\texttt{CodeXEmbed} and \texttt{PDG}) align with this aggregate well above chance ($0.709$ and $0.657$), confirming their ability to coarsely capture algorithmic differences in their respective spaces. Therefore,
we adopt a two-stage approach to obtain the algorithmically diverse solutions. As the dataset contains 1.6M correct C++ submissions after the primary filtration, we cannot rely on LLM-based pairwise measurements to obtain algorithmically diverse solutions due to cost constraints. Thus, we first use the two lightweight similarity measures \texttt{CodeXEmbed} and \texttt{PDG} to filter similar solutions, and then select the final diverse solution pairs using the best LLM judges.

For the first stage, we filter solutions from human submissions (median: 185 solutions) to retain only those with higher algorithmic diversity, as captured by the lightweight metrics \texttt{CodeXEmbed} and \texttt{PDG}. In each metric space, we kept instances using a medoid-seeded farthest-first traversal strategy to capture distant solutions. 
We set the number of solutions filtered per strategy based on the covering-radius curve, stopping when an additional solution pick reduced the normalized radius by less than 2 percentage points. We ended up sampling 10 solutions from the \texttt{PDG} space, 6 from the \texttt{CodeXEmbed} space, and 5 uniform random solutions as controls to ensure appropriate representatives for each solution. As the samples were done independently, this filtration stage yielded a median of 19 solutions per instance.


For the subsequent LLM-based filtering step, we shortlisted the disjoint top-3 pairs of solutions, ranked in ascending order of similarity computed as the mean \texttt{CodeXEmbed} and \texttt{PDG} similarity scores. We used the cheaper \texttt{Gemini-3.1-flash-lite} and \texttt{gpt-5.4-mini} to establish algorithmic dissimilarity between the pairs, using \texttt{Claude-Haiku-4.5} to break the ties. 

\input{Tables/data_stats}

\textbf{Ontology-driven Program Classification for Interpretable Labels.}
To collect labels for the program description, we propose a three-layer ontology to describe a program's algorithmic paradigm, the techniques used in the problem solving, and the data structures used. The ontology spans $13$ paradigms, $64$ techniques, and $22$ data structures. We used \texttt{Gemini-3.1-flash-lite} for this ontology labeling, as it had the highest agreement with the other six proprietary models we considered (see Appendix~\ref{app:subsec:ontology}). We provide the distribution of the ontology labels in Figure~\ref{fig:ontology-distribution}. 

\textbf{Final dataset statistics.} We split our dataset into training, validation and test split with stratified sampling based on the programming paradigms and the average length of candidate C++ solutions (see Table~\ref{tab:dataset_splits}). We additionally classify each solution pair's algorithmic diversity into four categories depending on their ontology label overlap- \textit{strong}: different programming paradigms, \textit{medium}: same programming paradigms but different techniques, \textit{weak}: same programming paradigm and technique, but different data structure, and \textit{identical}: same ontology for both solutions. We provide the examples for each classification type in the Figures~\ref{fig:example-strong},~\ref{fig:example-medium},~\ref{fig:example-weak} and~\ref{fig:example-identical}.


%% file: Tables/data_stats.tex
\begin{table*}[t]
\centering
\small
\setlength{\tabcolsep}{4pt}
\caption{Dataset statistics of the algorithmic diversity benchmark.}
\label{tab:dataset_splits}
\resizebox{0.85\linewidth}{!}{%
\begin{tabular}{lcccccccc}
\toprule
& & & \multicolumn{2}{c}{\textbf{Human solutions}} & \multicolumn{4}{c}{\textbf{Ontology-based algorithmic diversity}} \\
\cmidrule(lr){4-5}\cmidrule(lr){6-9}
Split & \textit{instances} & \textit{C++ sol.} & \textit{total} & \textit{median} &
\textit{strong} & \textit{medium} & \textit{weak} & \textit{identical} \\
\midrule
train & 5,623 & 11,246 & 148,242 & 25 & 2,429 & 1,463 & 1,229 & 502 \\
val   & 1,000 &  2,000 &  25,915 & 24 &   419 &   257 &   231 &  93 \\
test  & 1,000 &  2,000 &  26,145 & 25 &   428 &   280 &   212 &  80 \\
\midrule
all   & 7,623 & 15,246 & 200,302 & 25 & 3,276 & 2,000 & 1,672 & 675 \\
\bottomrule
\end{tabular}}
\end{table*}

%% file: Sections/5_exp_setting.tex
\section{Experimental Settings} \label{sec:exps}
Section~\ref{sec:prob-def} defines the problem, covering two types of algorithmic diversity: population-level and algorithm coverage over human solutions. Section~\ref{subsec:metric} 
describes the evaluation metrics and 
Section~\ref{subsec:baselines} presents 
the models and generation strategies used to answer our 3 research questions. 


\input{Sections/3_prob_def}

\subsection{Evaluation Metrics}\label{subsec:metric}

\noindent\underline{\textbf{$\mathrm{\mathbf{Vendi}}_{\texttt{CodeX-SFT}}$}: algorithmic diversity of a set of solutions.} To measure the number of algorithmically diverse solutions in a set of solutions, we distilled the \texttt{CodeXEmbed} embedding model~\cite{liu2024codexembed} on solution pairs that were labeled with the three proprietary LLM ensemble in the second stage filtering (Section~\ref{sec:dataset}). The training data comprised 14,195 solution pairs from the training and validation instances, and achieved improved alignment with a six-LLM ensemble (0.727 AUC compared to 0.686 AUC on the base model). We also validate the metric's quality on a small, human-curated algorithmic diversity dataset from~\citet{zhang2026rethinking}. We show that even for out-of-distribution Python programs, the trained model outperforms the base model in distinguishing behaviorally distinct solutions (see Appendix~\ref{appx:trained-embedding}). As a small model with moderately high alignment with frontier models, we used this similarity measure as our primary algorithmic similarity metric and refer to the trained model as \texttt{CodeX-SFT}. 
We adopt the Vendi score~\cite{friedman2022vendi} to evaluate the diversity of a population of solutions. For each instance with a set of $n$ solutions, we embed them with \texttt{CodeX-SFT}, and L2-normalize them to obtain the cosine Gram matrix. The Vendi score is then obtained over this matrix, with outputs ranging between $1$ (all solutions are identical) and $n$ (all solutions are mutually dissimilar) as follows:
\begin{equation} \label{eq:vendi}
\mathrm{Vendi}_{\texttt{CodeX-SFT}} \;=\; \exp\!\left(-\sum_i \lambda_i \log \lambda_i\right)
\end{equation}
where $\lambda_1 \dots \lambda_n$ are the eigenvalues of the cosine Gram matrix. 

\noindent\underline{$\mathrm{\mathbf{AlgoCov}}$: algorithmic coverage of representative human solutions.} To evaluate whether a set of solutions $S = \{s_1, s_2, \dots, s_N\}$ is able to express the two representative human solutions (referred to as A and B), we propose the metric $\mathrm{Coverage}$. We first obtain the ontology-labels for the solutions using the \texttt{Gemini-flash-lite-3.1} model $\{D_{s_1}, \dots, D_{s_N}\}$ and $\{D_A, D_B\}$. We compute the solution set's algorithmic coverage over the two solutions as follows: 
\begin{equation}\label{eq:algocov}
\begin{split}
    \textrm{Match}_X(s_i) = \mathbbm{1}\!\left[ D_{s_i} \cap D_X \neq \emptyset \right], \quad X \in \{A, B\}\\
    \textrm{AlgoCov}(S) = \max_{s_i \in S} \textrm{Match}_A(s_i) \cdot \max_{s_j \in S} \textrm{Match}_B(s_j)    
\end{split}
\end{equation}






\subsection{Code generation models and generation strategies} \label{subsec:baselines}

To evaluate uncontrolled diversity of LLM-generated solutions, we tested 24 code generation models: 2 closed-source proprietary models \texttt{Gemini-3.1-Flash-Lite} and \texttt{Gemini-3.8-Flash} and 22 open-source models \texttt{Gemma-4} (\texttt{E4B, 12B, 26B, 31B})~\citep{team2026gemma}, \texttt{codegemma-7b-it}~\citep{team2024codegemma},  \texttt{Phi-4}~\citep{abdin2024phi}, \texttt{CodeQwen-1.5-7B}~\citep{bai2023qwen}, \texttt{Qwen2.5} (\texttt{0.5B, 7B, Coder-7B})~\citep{qwen25}, \texttt{Qwen3-Coder} (\texttt{30B} and \texttt{32B})~\citep{yang2025qwen3}, \texttt{Deepseek-V2-Lite}~\citep{liu2024deepseek}, \texttt{Codestral-22B}, \texttt{Starcoder2-15B}~\citep{lozhkov2024starcoder}, \texttt{Mistral-7B-v0.3}~\citep{jiang2023mistral}, \texttt{Mistral-Nemo}, \texttt{Llama} (\texttt{3.2-3B}, \texttt{3.1-8B})~\citep{grattafiori2024llama} and \texttt{CodeLlama} (\texttt{7B}, \texttt{13B}, \texttt{34B})~\citep{roziere2023code}. We then select three open-weight models (\texttt{Phi-14B}, \texttt{Gemma-4-E4B} and \texttt{Gemma-4-31B}) for our diverse code generation experiments, 
based on high pass rates with diverse models on the uncontrolled generations ($31.4\%$, $44.7\%$ and $76.8\%$ respectively; Table~\ref{tab:results_raw_runs}). The diverse code generation strategies include temperature variations in the uncontrolled generation prompt, \textit{verbalized} sampling~\citep{zhang2025verbalized}, and \textit{iterative} constrained generation~\citep{ye2025assessing}, in which the model was prompted to generate a valid solution different from what it had already produced.  
These uncontrolled and controlled generation experiments, evaluated with $\mathrm{Vendi}_{\texttt{CodeX-SFT}}$, help us answer our first two research questions, \textit{how diverse are single model generated solutions} (\textbf{RQ1}; Section~\ref{sec:rq1}), and \textit{how algorithmically distinct solutions from different model generated solutions are} (\textbf{RQ2}; Section~\ref{sec:rq2}). We also conduct steering experiments to examine whether \textit{models can generate solutions following the human strategies} (\textbf{RQ3}; Section~\ref{sec:rq3}). For this experiment, we provide the ontology labels from the reference solutions A and B (called ``steer (A+B)"), and evaluate coverage using $\mathrm{AlgoCov}$ to estimate the oracle performance. We provide the hyperparameters in Appendix~\ref{appx:hyperparams} and the prompts in Appendix~\ref{appx:prompts}.



%% file: Sections/3_prob_def.tex
\subsection{Problem Definition} \label{sec:prob-def}

Given a problem $P$, let $S^h_P = \{s^h_1, s^h_2, \dots, s^h_N\}$ be a population of $N$ correct human-written solutions to $P$. We assume an algorithmic diversity function $D: 2^{\mathcal{S}} \rightarrow \mathbb{R}$ that maps a population of programs to a scalar measuring the distinct underlying algorithms in $\mathcal{S}$. Then, a code generation model $\mathbf{M}$ that generates $N$ solutions $S^{\mathbf{M}}_P = \{s^{\mathbf{M}}_1, \dots, s^{\mathbf{M}}_N\}$, is considered as algorithmically diverse as humans, if $D(S^{\mathbf{M}}_P) \geq D(S^h_P)$ provided $D(S^h_P) \geq 1$ (\textit{i.e.,} more than one solution exists for this problem). Additionally, assuming a labeling function $\mathcal{A}: \mathcal{S} \rightarrow 2^{\mathcal{L}}$ that maps each program to a set of algorithmic approaches over a label space $\mathcal{L}$; a model $\mathbf{M}$ captures human algorithmic diversity on $P$ exactly when $\mathcal{A}(S^{\mathbf{M}}_P) = \mathcal{A}(S^h_P)$, a strictly stronger condition than $D(S^{\mathbf{M}}_P) \geq D(S^h_P)$.

%% file: Sections/6_results.tex
\section{Results and Discussion} \label{sec:results}

\begin{figure}
    \centering
    \includegraphics[width=0.65\linewidth]{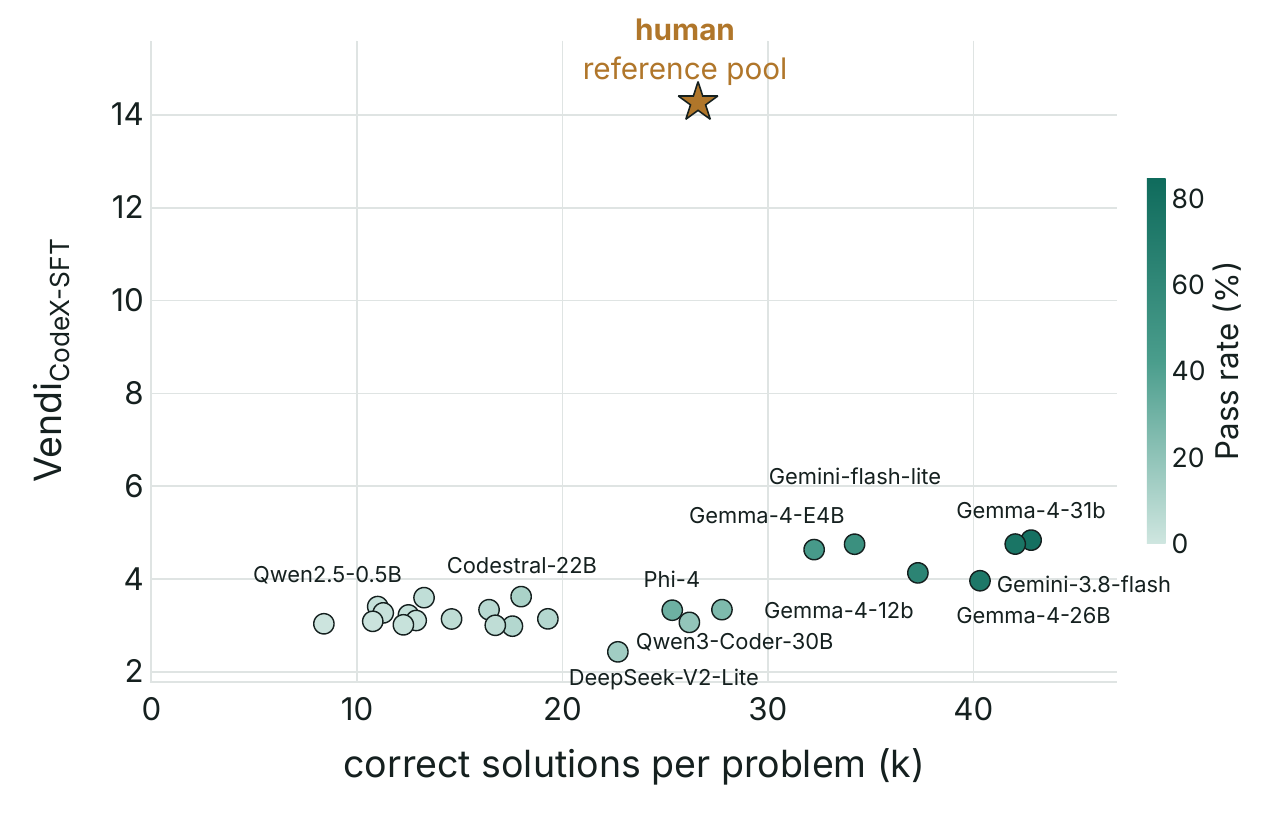}
    \caption{Algorithmic diversity ($\mathrm{Vendi}_{\texttt{CodeX-SFT}}$) of 50 uncontrolled LLM-generated solutions against human-written solutions for the test set problems (detailed results in Table \ref{tab:results_raw_runs} Appendix~\ref{appx:results}).} 
    \label{fig:vendi-results}
\end{figure}


\subsection{How diverse are LLM-generated solution?} \label{sec:rq1}

To answer \textbf{RQ1}, we generate 50 unconditioned solutions by zero-shot prompting the LLMs over the 1,000 test set instances. Figure~\ref{fig:vendi-results} illustrates the algorithmic diversity of 24 LLMs and the human solution diversity using the $\mathrm{Vendi}_{\texttt{CodeX-SFT}}$ metric (Eq.~\ref{eq:vendi}). We found that unconditioned generations do not yield very diverse solutions, even after sampling almost twice as many solutions as the available human reference pool. Compared to the $\mathrm{Vendi}$ score of 14.27 for human solutions, the most diverse LLM (\texttt{Gemini-flash-3.8}) achieved 4.84 $\mathrm{Vendi}$ score. 

\begin{figure}[t!]
    \centering
    \begin{subfigure}[t]{0.54\textwidth}
        \centering
        \includegraphics[width=\linewidth]{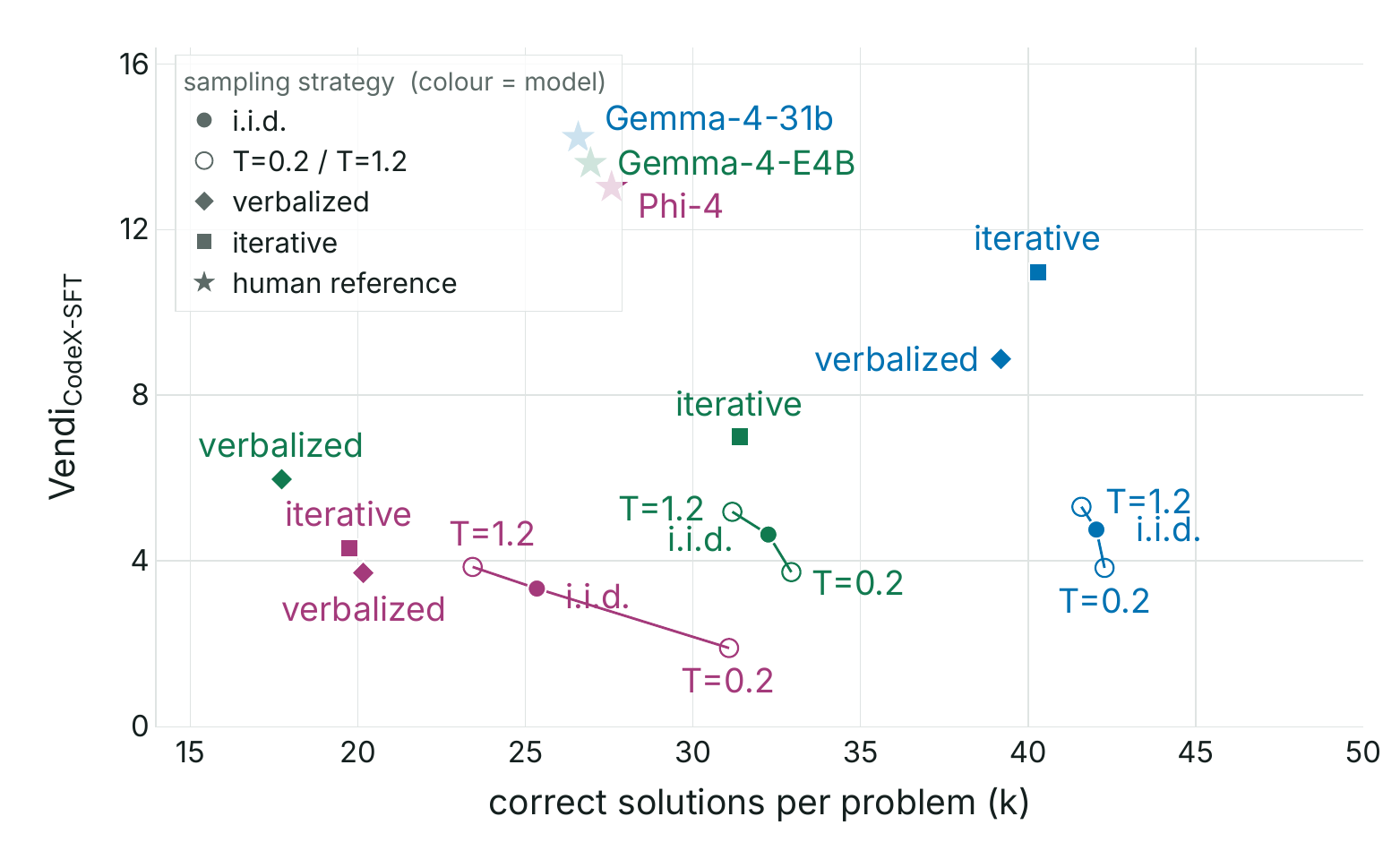}
    \end{subfigure}
    \begin{subfigure}[t]{0.45\textwidth}
        \centering
        \includegraphics[width=\linewidth]{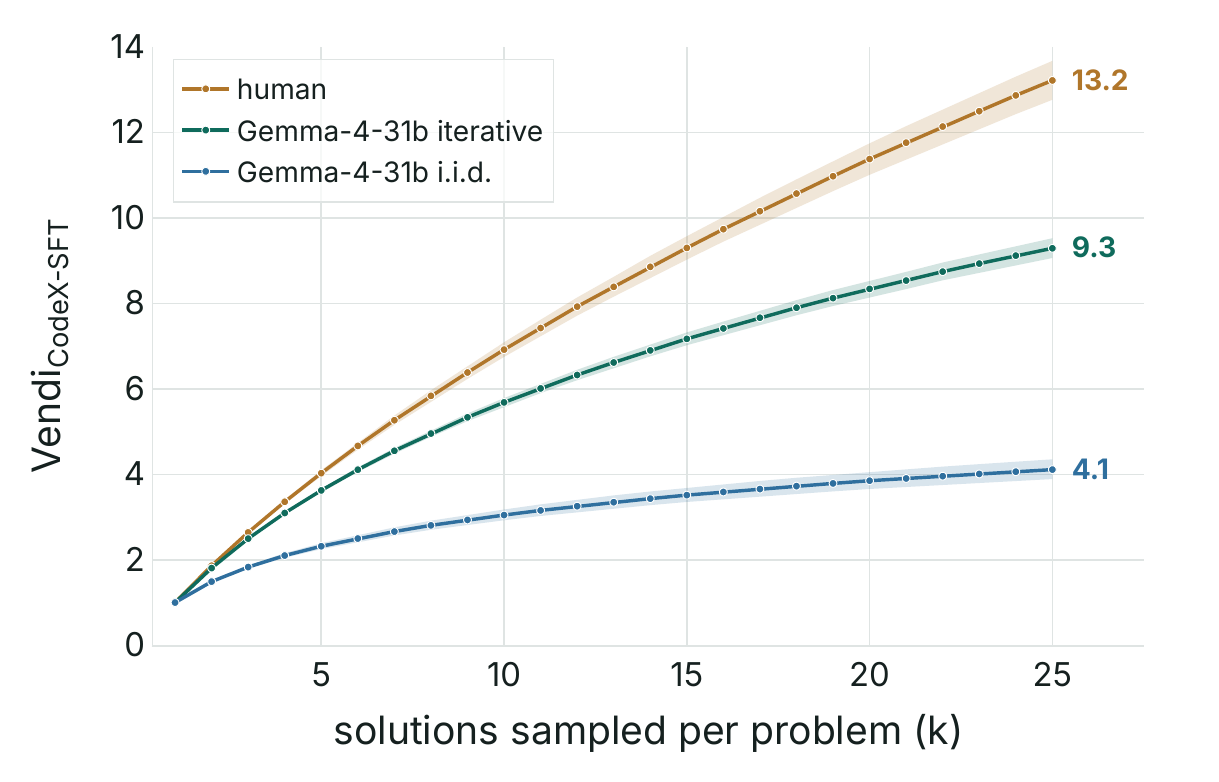}
    \end{subfigure}
    \caption{Algorithmic diversity ($\mathrm{Vendi}_{\texttt{CodeX-SFT}}$) of three diverse text generation strategies for three LLMs. (Left) temperature variation, \textit{verbalized} sampling, and \textit{iterative} constrained generation on the test set problems and the corresponding scores for human solutions for matched problems. (Right) increase in algorithmic diversity of human solutions and the best-performing LLM, \texttt{Gemma-4-31b}, for 412 problems where at least 25 correct solutions are available for both human and LLM solutions. T=0.8 is denoted as \textit{i.i.d.}, independent and identically distributed samples (same as the results in Figure~\ref{fig:vendi-results}). We report the human solutions corresponding to each model's success rate to enable a fair comparison across model generations. For the detailed results, refer to Table \ref{tab:steer_internal} in Appendix~\ref{appx:results}.} \label{fig:vendi-diversity}
\end{figure}

To further investigate the extent of LLM-generated diverse solutions, we selected three representative open-weight models \texttt{Gemma-4-31b}, \texttt{Gemma-4-E4b} and \texttt{Phi-4-14b} to inspect the impact of temperature variation, \textit{verbalized} sampling and \textit{iterative} generation (Section~\ref{subsec:baselines}). Figure~\ref{fig:vendi-diversity} (left) presents the results from this experiment. We found that increasing temperature increased the diversity of the correctly generated solutions, with a ${\sim}+1$ increase from T=0.2 to T=0.8 and a $+0.5$ increase from T=0.8 to T=1.2. However, increasing the temperature affected the models differently, with weaker models yielding fewer correct solutions than stronger models, which had higher pass rates. In comparison, the other two diverse output generation strategies were significantly more effective, with verbalized sampling increasing the diversity from $+0.38$ to $+4.13$ and iterative generation from $+0.98$ to $+6.22$. While these strategies outperform the zero-shot baseline, their higher $\mathrm{Vendi}$ score was partially attributable to having more samples than the human reference. For a fair comparison, we examined the best-performing model, \texttt{Gemma-4-31b} with iterative generation, against the human-written solutions. We selected the subset of problems (N=412) with at least 25 solutions for both populations. We found that while iterative generation did significantly improve the solution diversity compared to the zero-shot (\textit{i.i.d.}) baseline, it also steadily remained below the human solutions, achieving 70.5\% of the $\mathrm{Vendi}$ score for the sample size of 25 solutions on this subset (Figure~\ref{fig:vendi-diversity} right).

\textbf{Discussion.} These results indicate that algorithmic diversity is not a property that emerges from sampling alone, irrespective of the number of \textit{i.i.d.} samples generated. The most diverse model barely reaches a third of the human $\mathrm{Vendi}$ score. Simple strategies like temperature increases yield diminishing returns while degrading correctness in weaker models. Prompting strategies that elicit variation are considerably more effective, suggesting that the diversity deficit stems from how the solutions are sampled rather than the model choice. The strongest single-model approach still captures only 70\% of human diversity at a matched sample size, highlighting the limitations of LLMs in eliciting human-level algorithmic variety.

\subsection{Do Models Converge on the Same Algorithms?} \label{sec:rq2}

\begin{figure}[t!]
    \centering
    \begin{subfigure}[t]{0.4\textwidth}
        \centering
        \includegraphics[width=\linewidth]{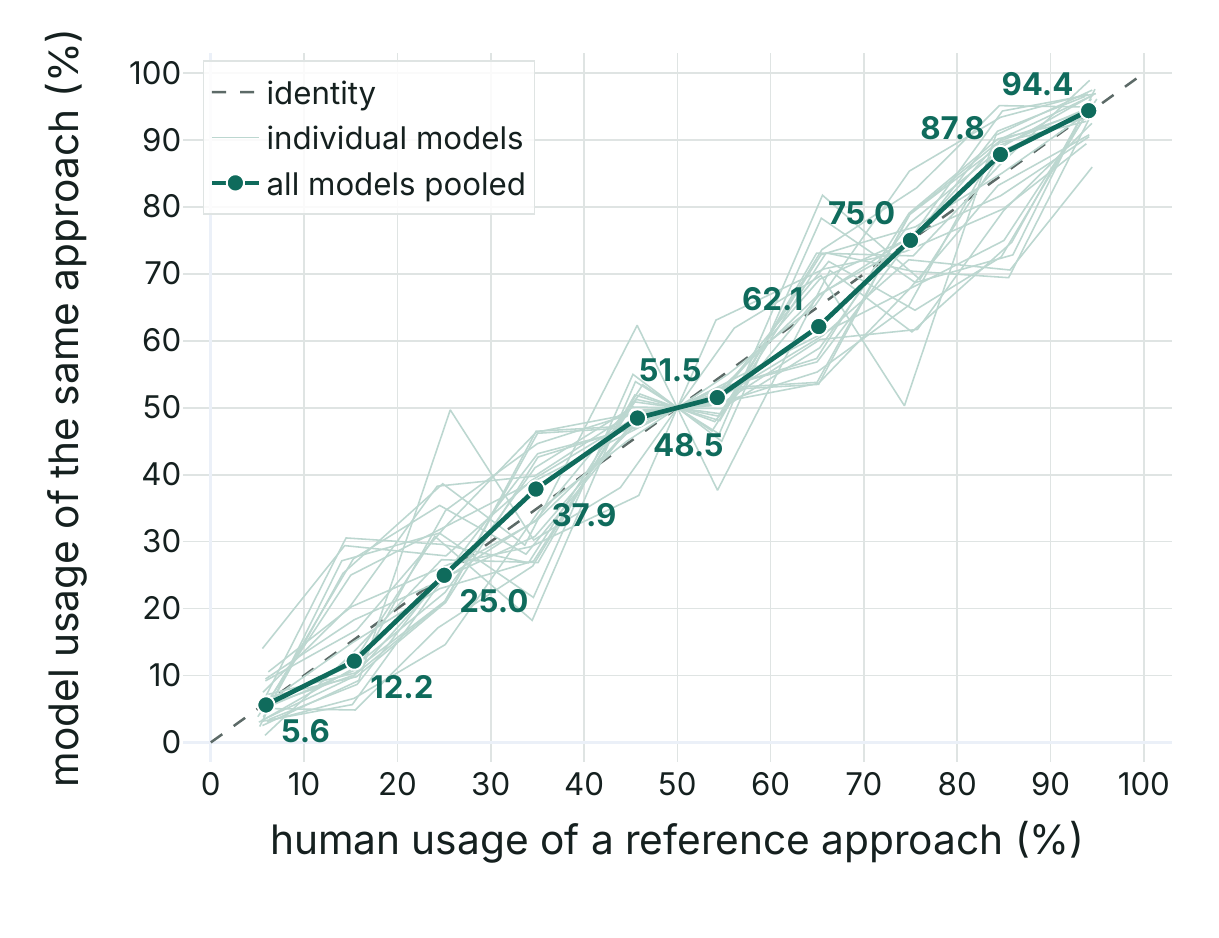}
    \end{subfigure}
    \begin{subfigure}[t]{0.5\textwidth}
        \centering
        \includegraphics[width=\linewidth]{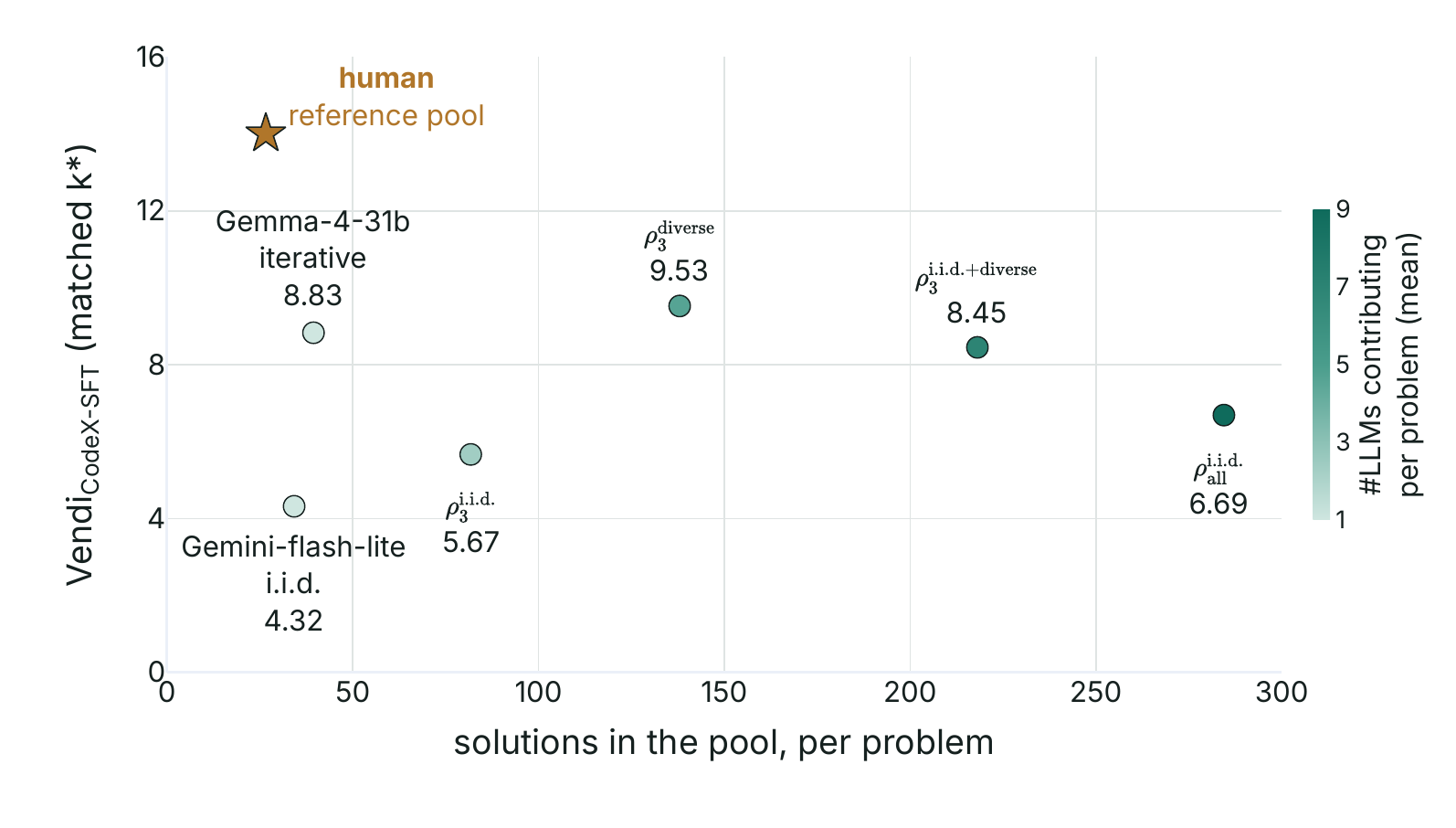}
    \end{subfigure}
    \caption{(left) LLM-generated solution's preference for using one of the two reference approaches vs. the popularity of the same approach in human-written solutions. Alignments for model and human preferences were obtained using the cosine similarities obtained from the \texttt{CodeX-SFT} embedding model. (right) Algorithmic diversity of ensemble of LLM-generated programs $\rho$ at the matched $k^*$ to the human reference solutions along with the most diverse LLMs for \textit{i.i.d.} setting and the \textit{diverse} code generation settings. $\rho_{all}$ denotes all 24 LLMs used in this work, $\rho_{3}$ denotes the three open-source models used for diverse code generation experiments (Figure~\ref{fig:vendi-diversity} left).} \label{fig:model-ensemble}
\end{figure}

To answer \textbf{RQ2}, we investigate the algorithmic preference of LLM-generated solutions juxtaposed against the popularity of the approach within the human reference solutions. We measured algorithmic similarity of the solution population against the two representative solutions using the trained \texttt{CodeX-SFT} with a 0.23 cosine distance threshold\footnote{
The threshold was obtained by calibrating the algorithmic similarity metric in the \texttt{CodeX-SFT} metric space using two representative human solutions (A and B) from 150 randomly selected instances in the dataset. The cutoff of the 10th percentile was chosen to separate the \textit{matched} and \textit{novel} solution pairs on the known algorithmically distinct pairs. This conservative threshold helped establish the boundary of algorithmically similar solutions for the experiment.}. Our analysis (Figure~\ref{fig:model-ensemble} left) indicated a clear affinity of LLM-generated solutions to align with the more popular representative solution (mean Pearson's $r=+0.695$). 

Additionally, we computed the inter-model solution diversity by sampling solutions from different unions of LLM-generated solutions (Figure~\ref{fig:model-ensemble} right). Since the $\mathrm{Vendi}$ score depends on the number of samples, we evaluated the diversity for a subset of solutions for all compared solutions ($k^* = min(k_{human}, k_{LLM})$). We found that sampling from heterogeneous model generations does increase the solution diversity ($\mathrm{Vendi}(\rho_{all}^{i.i.d.}) = 6.69$) compared to the most diverse individual LLM ($\mathrm{Vendi}(\texttt{Gemini-flash-3.8}) = 4.84$); it still sits far below the diverse code generation strategies ($\mathrm{Vendi}(\texttt{Gemma-4-31b}_{iterative}) = 8.83$). Similarly, sampling from an ensemble of the three models prompted to generate diverse solutions also increased the diversity ($\mathrm{Vendi}(\rho_{3}^{diverse}) = 9.53$); however, the gain was significantly less for the computational cost incurred to obtain the additional solutions compared to a single model. 

\textbf{Discussion.} Our results corroborate the findings from \citet{lee2025diversely}, who show that merging solutions from heterogeneous models increases diversity. While we found that a union of multiple models increases diversity over a single model (from 4.32 to 6.69), it is not a substitute for diverse code generation. Combining multiple, diversely generated outputs further increases diversity (from 8.83 to 9.53); the additional +0.70 gain over the best single-model approach is marginal compared to the threefold increase in generation costs. This finding, along with the strong alignment between model preferences and the more popular human approach ($r=+0.695$), suggests that heterogeneous models share an algorithmic prior that cannot be overcome by ensembling the model outputs. 

\subsection{Can LLM-generated solutions follow human solution strategies?} \label{sec:rq3}

To answer \textbf{RQ3}, we use our algorithmic diversity benchmark to investigate whether LLMs are capable of capturing the two 
representative human solutions using $\textrm{AlgoCov}$ (Eq.~\ref{eq:algocov}). To specifically measure the coverage of algorithms expressed in the reference solutions (depicted as A and B), we limit this investigation to a subset of 484 problems in the test set with mutually exclusive solutions in \textit{strong} (274), \textit{medium} (104) and \textit{weak} (106) overlap granularities. In addition to the diverse generation strategies discussed earlier, we include the ``steer (A+B)" approach that explicitly provides the ontology labels from the reference solutions to steer the model to generate the corresponding solutions (prompt in Figure~\ref{fig:prompt-steer-ab}). This oracle-style baseline serves as an upper bound on how well LLMs can reproduce the reference solutions when given explicit guidance. To contextualize model performance using $\mathrm{AlgoCov}$, we additionally measure the coverage of non-reference human solutions for a grounded comparison.

Table~\ref{tab:coverage_tiers} shows the results from this study for a subset of 10 solutions per strategy. From the algorithmic coverage of unconditioned solutions (\textit{i.i.d.}) to the human coverage ceiling, we 
observe some clear trends across different solution generation strategies. For the best model (\texttt{Gemma-4-31b}), the trends are the most evident, with iterative generation strategy even surpassing the coverage compared to the explicit steer (A+B) baseline in two of the three ontology granularities\footnote{Verbalized sampling for the weaker models underperforms the unconditioned \textit{i.i.d.} baseline, not because it does not yield diverse solutions, but rather because these models do not generate enough correct samples. \texttt{Gemma-4-E4B} generated 3.3 average solutions in verbalized sampling, compared to 8.0 correct solutions in 10 generations in the \textit{i.i.d.} baseline.}. This indicates that the algorithmic strategies are--to some extent--latent rather than absent, and can be obtained without external steering. In comparison, the human-written solutions surpass all the generation strategies, with a significant gap in the weak setting where only the data structure differs (79.6\% compared to 23.3\%). 

We also evaluate the impact of explicit steering on the algorithmic strategies of the LLM-generated solution. We identified the model's default strategy by taking the aggregate intersection of the paradigm labels from half of the randomly selected \textit{i.i.d.} solutions with the reference solutions, yielding a default and a non-default approach. We found that specifying the model's default approach marginally increases its paradigm overlap from 91.5\% to 93.8\%. However, when asked to generate the non-default approach, the overlap increases from 15.4\% to 34.7\%. This gain was constant across the ontology granularities (+19.8, +16.6, +19.2 pp) for the non-default steered solutions.

\input{Tables/result_steer_external}

\textbf{Discussion.} These experiments suggest that the algorithmic strategies are latent rather than absent, as indicated by the iterative generation matching the oracle steer (A+B) in two granularities and steering toward a non-default paradigm more than doubles the overlap (from 15.4\% to 34.7\%). The gap in coverage of the default approach remains significantly higher than that of the non-default approach (93.8\% \textit{vs.} 34.7\%), indicating that, while steerable, LLMs still struggle to express the less-known strategy. In comparison, the human solutions surpass the coverage across all three granularities, with a large gap in the weak tier indicating that diversity in fine-grained data structure choices is harder to elicit than diversity in programming paradigms. We acknowledge that $\mathrm{AlgoCov}$, as instantiated here, assumes exactly two representative solutions and does not reward other valid strategies beyond the two solutions. The human results support this argument, with coverage of 55-80\% across granularities, indicating that there are more than two ways to solve the problems. While it could be feasible to extend our benchmark beyond 2 representative human solutions, each additional solution would need to be distinct from the existing ones, thereby compounding the number of similarities to be computed.
We argue that the human-grounded ceiling does help situate the model results better, and we leave the exploration beyond two reference solutions to 
future work.

%% file: Tables/result_steer_external.tex
\begin{table*}[t]
\centering
\small
\setlength{\tabcolsep}{5pt}
\caption{Algorithmic coverage ($\textrm{AlgoCov}$) of LLM-generated solutions and corresponding human-written solutions over the two human solutions in the final dataset (referred as A and B). We sampled 10 solutions for each strategy; limiting to 274 \textit{strong}, 104 \textit{medium} and 106 \textit{weak} instances where the ontology-labels are mutually exclusive for the respective classification dimension.} 
\label{tab:coverage_tiers}
\resizebox{0.7\linewidth}{!}{%
\begin{tabular}{lrccccc}
\toprule
Model & $n$ & \textit{i.i.d.} & verbalized & iterative & steer (A$+$B) & human \\
\midrule
\multicolumn{7}{l}{\textbf{Strong} --- targets differ in paradigm (274 problems)} \\
\texttt{Gemma-4-31b} & 245 & 28.2 {\scriptsize\textcolor{gray}{[22,33]}} & 50.2 {\scriptsize\textcolor{gray}{[44,57]}} & \textbf{60.8} {\scriptsize\textcolor{gray}{[54,67]}} & 46.6 {\scriptsize\textcolor{gray}{[41,52]}} & 70.0 {\scriptsize\textcolor{gray}{[66,75]}} \\
\texttt{Gemma-4-E4B} & 179 & 34.1 {\scriptsize\textcolor{gray}{[27,41]}} & 31.3 {\scriptsize\textcolor{gray}{[25,38]}} & 44.1 {\scriptsize\textcolor{gray}{[37,51]}} & \textbf{48.3} {\scriptsize\textcolor{gray}{[42,55]}} & 70.8 {\scriptsize\textcolor{gray}{[65,76]}} \\
\texttt{Phi-4} & 121 & 24.8 {\scriptsize\textcolor{gray}{[17,33]}} & \textbf{29.8} {\scriptsize\textcolor{gray}{[22,38]}} & 24.8 {\scriptsize\textcolor{gray}{[17,32]}} & 26.9 {\scriptsize\textcolor{gray}{[20,34]}} & 67.7 {\scriptsize\textcolor{gray}{[61,74]}} \\
\midrule
\multicolumn{7}{l}{\textbf{Medium} --- targets differ in technique (104 problems)} \\
\texttt{Gemma-4-31b} & 87 & 21.8 {\scriptsize\textcolor{gray}{[14,31]}} & 32.2 {\scriptsize\textcolor{gray}{[23,43]}} & 37.9 {\scriptsize\textcolor{gray}{[28,48]}} & \textbf{40.9} {\scriptsize\textcolor{gray}{[32,50]}} & 59.0 {\scriptsize\textcolor{gray}{[50,68]}} \\
\texttt{Gemma-4-E4B} & 49 & 16.3 {\scriptsize\textcolor{gray}{[6,27]}} & 14.3 {\scriptsize\textcolor{gray}{[6,24]}} & 32.7 {\scriptsize\textcolor{gray}{[20,47]}} & \textbf{36.8} {\scriptsize\textcolor{gray}{[25,49]}} & 56.8 {\scriptsize\textcolor{gray}{[44,68]}} \\
\texttt{Phi-4} & 26 & 11.5 {\scriptsize\textcolor{gray}{[0,23]}} & 15.4 {\scriptsize\textcolor{gray}{[4,31]}} & 7.7 {\scriptsize\textcolor{gray}{[0,19]}} & \textbf{26.6} {\scriptsize\textcolor{gray}{[12,43]}} & 54.9 {\scriptsize\textcolor{gray}{[38,71]}} \\
\midrule
\multicolumn{7}{l}{\textbf{Weak} --- targets differ in data structure (106 problems)} \\
\texttt{Gemma-4-31b} & 90 & 10.0 {\scriptsize\textcolor{gray}{[4,17]}} & 17.8 {\scriptsize\textcolor{gray}{[10,26]}} & \textbf{23.3} {\scriptsize\textcolor{gray}{[14,32]}} & 15.5 {\scriptsize\textcolor{gray}{[8,23]}} & 79.6 {\scriptsize\textcolor{gray}{[72,87]}} \\
\texttt{Gemma-4-E4B} & 71 & 5.6 {\scriptsize\textcolor{gray}{[1,11]}} & 4.2 {\scriptsize\textcolor{gray}{[0,8]}} & 7.0 {\scriptsize\textcolor{gray}{[1,13]}} & \textbf{10.1} {\scriptsize\textcolor{gray}{[4,17]}} & 77.8 {\scriptsize\textcolor{gray}{[69,86]}} \\
\texttt{Phi-4} & 59 & 5.1 {\scriptsize\textcolor{gray}{[0,12]}} & 3.4 {\scriptsize\textcolor{gray}{[0,8]}} & 5.1 {\scriptsize\textcolor{gray}{[0,12]}} & \textbf{8.2} {\scriptsize\textcolor{gray}{[3,15]}} & 76.9 {\scriptsize\textcolor{gray}{[67,86]}} \\
\bottomrule
\end{tabular}}
\end{table*}

%% file: Sections/8_conclusion.tex
\section{Conclusion} \label{sec:conclusion}

We introduced a fine-grained algorithmic diversity benchmark for code generation ($\texttt{DivAlgo}$) and presented a 
a distilled lightweight embedding model with LLM-ensemble annotations to practically measure population-level diversity of model-generated solutions at scale. Our extensive examination across multiple models and generation strategies reveals that unconditioned sampling recovers only a fraction of the algorithmic variety present in human solution pools, and that models converge on the more popular human approach. Prompting the models to generate diverse solutions closes some of this gap, unlike the minor gains from ensembling heterogeneous models. While steering toward a particular strategy helps improve the coverage of model-generated solutions, it still falls below the human ceiling across all three granularities. We conclude that algorithmic diversity is indeed a latent property of current LLMs, but they fail to secure the full variety of human solutions.

%% file: Sections/appx_a_data.tex
\section{Additional Dataset Construction details} \label{appx:dataset}

\subsection{Preliminary investigation of algorithmic similarity metrics}\label{subsec:preliminary}

\begin{figure}[h!]
    \centering
    \begin{subfigure}[b]{0.45\textwidth}
        \centering
        \includegraphics[width=\textwidth]{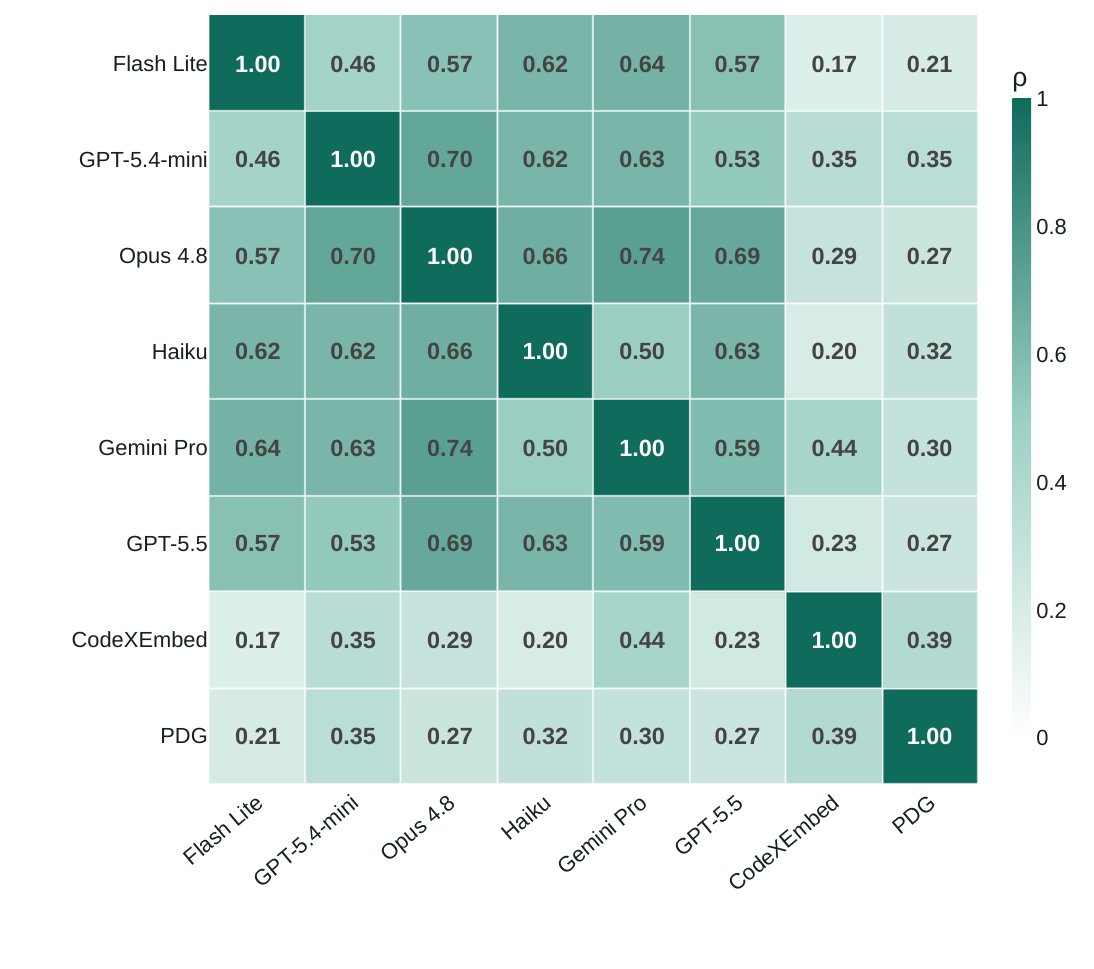}
    \end{subfigure}
    ~
    \begin{subfigure}[b]{0.45\textwidth}
        \centering
        \includegraphics[width=\textwidth]{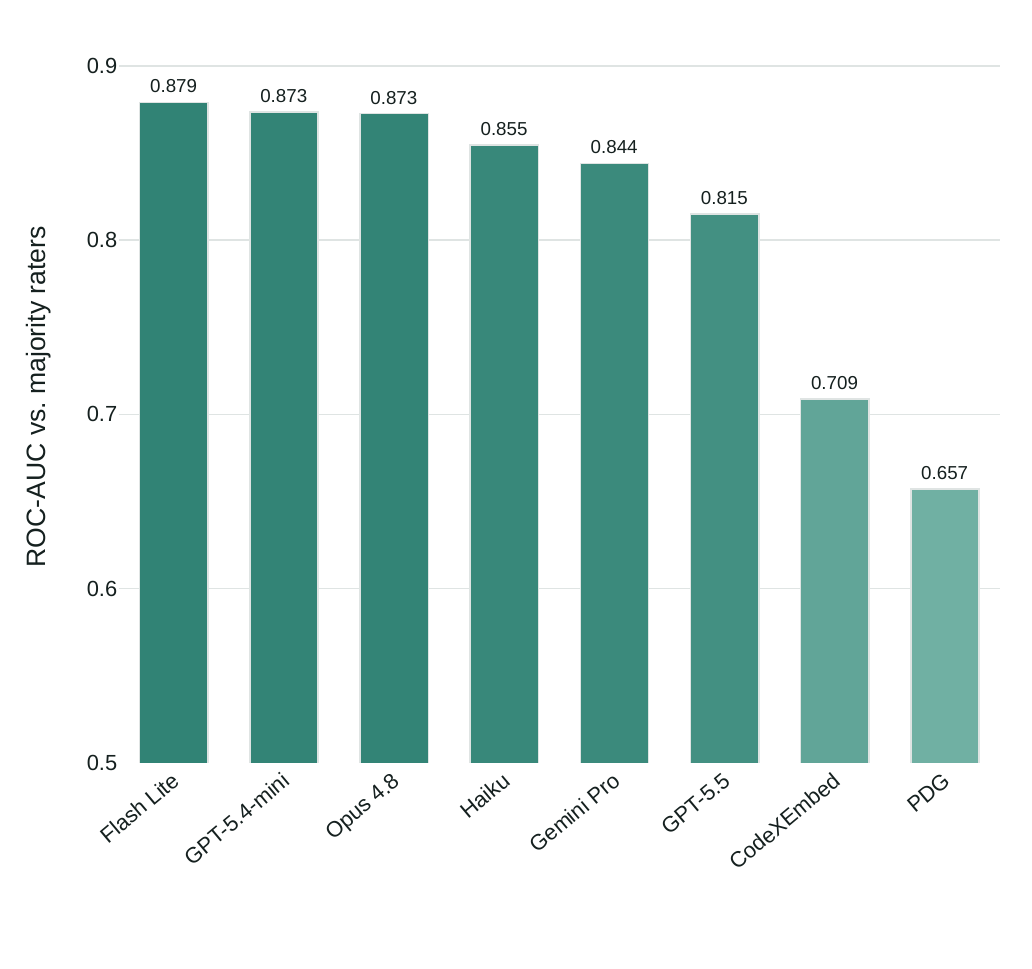}
    \end{subfigure}
    \caption{Algorithmic similarity metric agreements on 100 uniformly random solution pairs. (Left) Rank-based Spearman's $\rho$ and (right) ROC-AUC against the majority of metrics; for LLM-judges, the majority of the other 5 LLMs is used, 
    for \texttt{CodeXEmbed} and \texttt{PDG} all six LLM-judges are used.} 
    \label{fig:similarity-corr}
\end{figure}

To build \texttt{DivAlgo}, we need to identify algorithmically different solutions among the human-provided solutions. While obtaining human annotations for this task would be the best approach to obtain the gold-labels, a recent human-annotation study \citep{lee2025diversely} has shown that human correlation dropped from $\kappa = 0.58$ to $0.25$ when the edit distance increased from $100 < x \leq 500$ to $x \geq 500$. For the \texttt{CodeContests+} dataset, the average distance between the two C++ solutions is $\geq 1,800$, indicating the difficulty of obtaining high-quality human annotations, particularly at scale.  

Thus, we propose a systematic investigation of automatic metrics for algorithmic similarity and a two-stage diversity selection process (Section \ref{sec:dataset}) to build \texttt{DivAlgo}.  
To measure algorithmic similarity, we focus on  
similarity metrics that go beyond 
superficial syntactic attributes (such as variable names): 
1) program dependency graph (\texttt{PDG}) that captures the control and data dependencies for a program\footnote{Post-hoc edits were done on the program dependency graphs to remove several stylistic attributes (see Appendix~\ref{appx:hyperparams} for further details).}, 2) a code embedding model, 
\texttt{CodeXEmbed-2b} \cite{liu2024codexembed}, 
which was the best model on the code retrieval (CoIR) leaderboard~\cite{li2025coir} in July 2026, and 3) six proprietary LLMs (\texttt{Gemini-3.1-pro}, \texttt{Gemini-3.1-flash-lite}, \texttt{GPT-5.5}, \texttt{GPT-5.4-mini}, \texttt{Claude-opus-4.8} and \texttt{Claude-Haiku-4.5}~\cite{}) used as LLM-as-judge. The models were prompted to give the similarity classification out of four choices: \textit{identical}, \textit{variation},  \textit{distinct} and \textit{very distinct} (prompt in Appendix~\ref{appx:prompts}).
Figure~\ref{fig:similarity-corr} illustrates the correlations between the different similarity metrics for 100 uniformly sampled solution pairs from the \texttt{CodeContests+} dataset. We observed that while LLM-judges moderately correlated with each other on the rank-wise Spearman's $\rho$ (${\sim} 0.46{-}0.74$), they had a very high AUC (${\sim} 0.85$). This indicates that while LLMs disagreed with each other on the thresholds for what constitutes algorithmically similar, their ordering of programs is well aligned. Comparatively, the other two similarity measures also aligned moderately with the LLM-judges; AUC for \texttt{CodeXEmbed} was $0.709$, and \texttt{PDG} was $0.657$.


\begin{figure}[ht]
    \centering
    \includegraphics[width=0.65\linewidth]{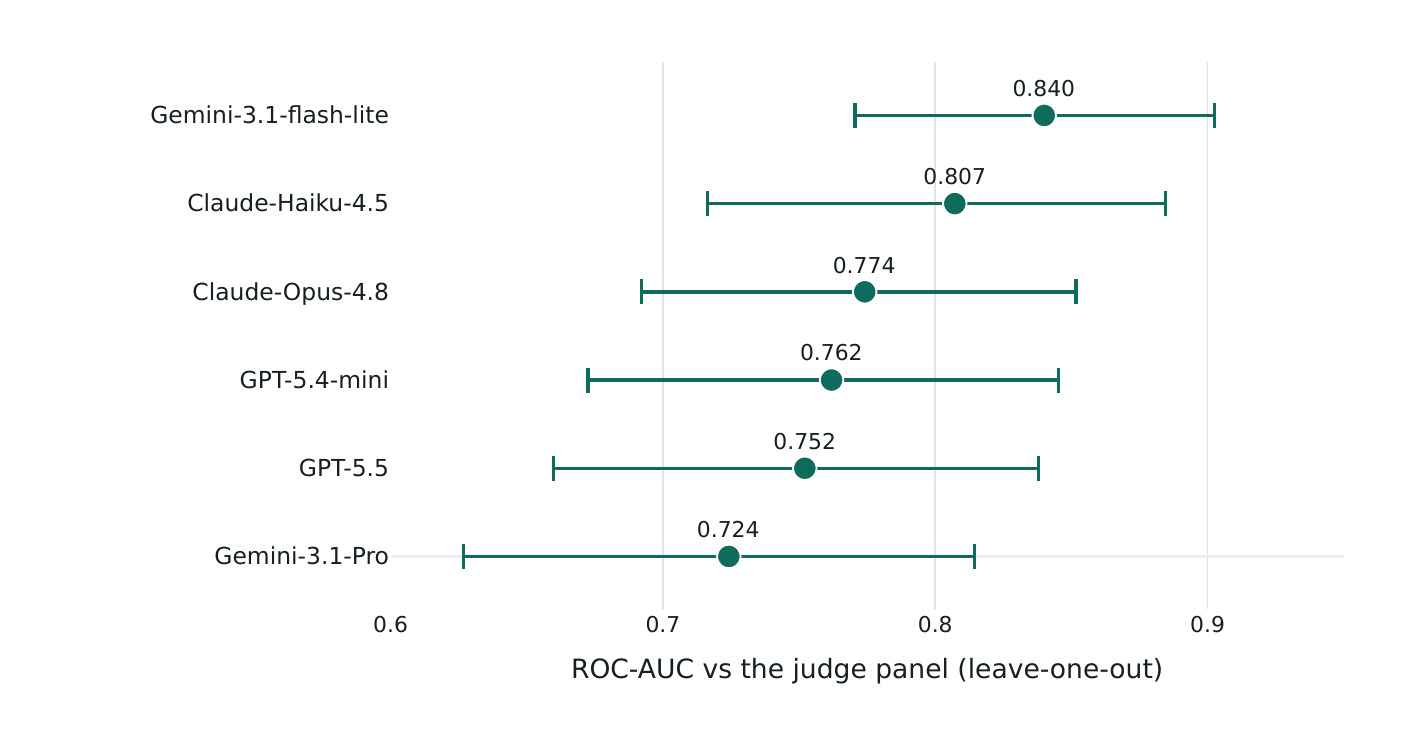}
    \caption{Leave-one-out agreement between six proprietary LLMs for the three-level ontology labeling task.}
    \label{fig:ontology-labeller}
\end{figure}

\subsection{Selecting the Program Classification LLM} \label{app:subsec:ontology}

We used an LLM-based classifier to label each programming solution with a three-level algorithm ontology that details the programming paradigm (13), technique (64), and data structure (22) used by the solution. To determine the best LLM for this task, we investigated six proprietary LLMs and found \texttt{Gemini-flash-lite-3.1} to have the highest agreement with the other five LLMs, as well as being the cheapest (see Figure~\ref{fig:ontology-labeller}). We therefore used it to obtain the ontology-based algorithm labels for the 15,246 C++ solutions in the final dataset. We present the distribution of the ontology-level-specific labels in Figure~\ref{fig:ontology-distribution}. While the model additionally provides labels beyond the ontology, we discard them and limit our classification to the predefined ontology labels. 

%% file: Sections/appx_b_metric.tex
\section{Calibrating \texttt{CodeXEmbed} as an evaluation metric} \label{appx:trained-embedding}

To select the best evaluation metric that was both open-source and efficient to run, we first constructed an evaluation subset from our dataset. The subset comprised 500 random solution pairs and 200 solution pairs from the primary filtration stage using \texttt{PDG} and \texttt{CodeXEmbed} similarities (refer to Section \ref{sec:dataset}).
The subset was intentionally designed: the random pairs calibrate the metric to the dataset's natural similarity distribution, while the filtration-stage pairs oversample the more difficult near-threshold regions that random sampling might not adequately represent. For each solution pair corresponding to a problem, we used six closed-source models (\texttt{Gemini-3.1-pro}, \texttt{Gemini-3.1-flash-lite}, \texttt{GPT-5.5}, \texttt{GPT-5.4-mini}, \texttt{Claude-opus-4.8} and \texttt{Claude-Haiku-4.5}~\cite{}) to obtain the ground-truth label on a four-point ordinal scale (0 - identical, 1 - variation, 2 - distinct, 3 - very distinct). The majority score from these six models was selected as the ground-truth rating for each instance, and we evaluated our lightweight metrics using the area under the ROC curve to evaluate their alignment.

\begin{figure}
    \centering
    \includegraphics[width=0.85\linewidth]{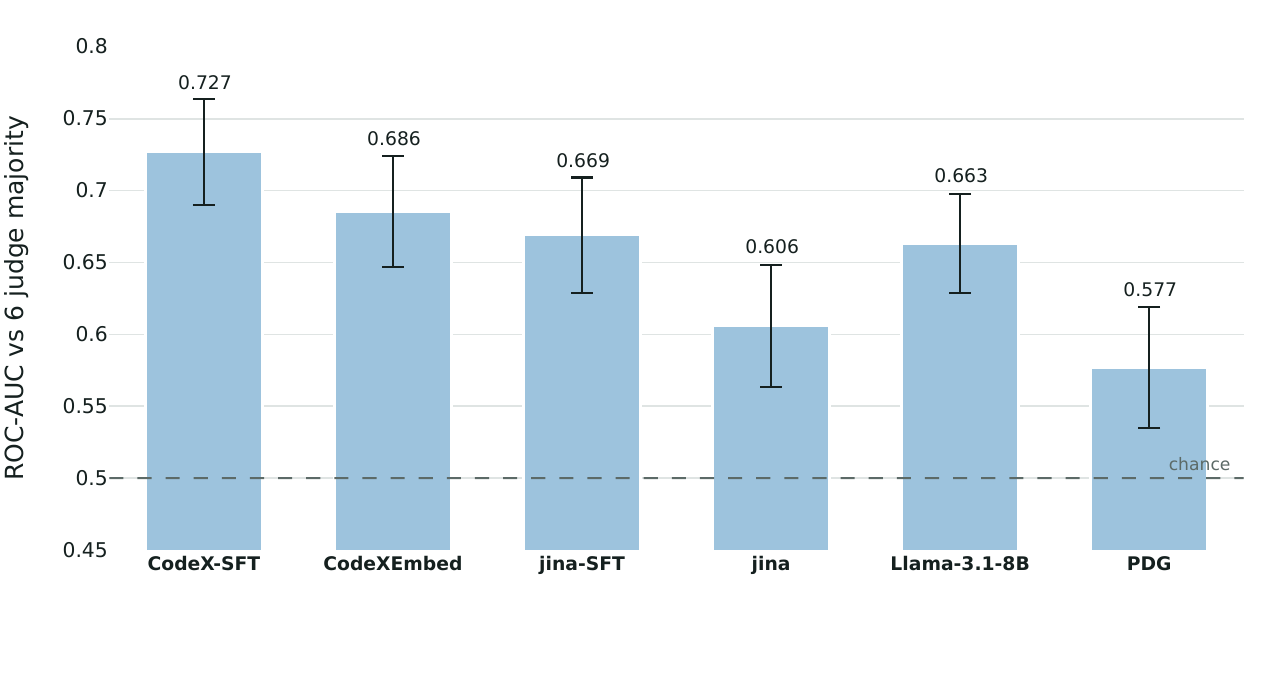}
    \caption{Evaluation of the algorithmic similarity measurement metrics on 700 instances, evaluated against the aggregated majority ground truth from six proprietary models. The confidence interval is obtained via percentile bootstrap over the solution pairs (repeated 3,000 times).}
    \label{fig:metric-eval}
\end{figure}

We evaluated various similarity metrics, from code embeddings \texttt{CodeXEmbed}~\cite{} and \texttt{jina}~\cite{}, code-graph based \texttt{PDG}, and LLM-as-judge using \texttt{Llama-3.1-8B} following~\cite{lee2025diversely}. We additionally fine-tuned the two code embedding models, distilling knowledge from the LLMs used to select the most distinct solutions during dataset curation (Section~\ref{sec:dataset}). We henceforth refer to these distilled embedding models as \texttt{CodeX-SFT} and \texttt{jina-SFT}. We trained the models on the labeled data we obtained from LLM-ensemble annotations during the second filtration stage (Section~\ref{sec:dataset}), excluding the 700 instances from the evaluation subset. This training dataset comprised 12,649 solution pairs from training and 1,546 from the validation split. The output distribution of the majority-voted labels obtained from the three proprietary LLM ensemble comprised 4,465 \textit{identical}, 2,880 \textit{variation}, 4,318 \textit{distinct}, and 2,532 \textit{very distinct} solution pairs. The target labels followed the same 4-way ordinal responses as the LLM-judge responses. We used mean-squared error as the loss function for this SFT training, training for 2 epochs with a batch size of 64, a $10^{-5}$ learning rate with 5\% warmup, and 8-bit AdamW.

We provide the meta-evaluation of similarity measures in Figure~\ref{fig:metric-eval}. All the evaluated metrics performed better than chance, but the only structural measure, \texttt{PDG}, showed the weakest alignment with the six-judge ensemble, achieving an ROC of 0.577. \texttt{Llama-3.1-8B} performs similarly to the \texttt{CodeXEmbed} model, although it has more than three times the number of parameters. For both models, fine-tuning improved embedding performance, with the best-performing metric being \texttt{CodeX-SFT}. We therefore used this metric as the basis for our algorithmic diversity evaluation, computing the Vendi Score across pairwise \texttt{CodeX-SFT} similarities. We also provide the self-diversity scores of \texttt{PDG} and \texttt{CodeX-SFT} to provide complementary insights into the diversity of a population of solutions, measured as the mean pairwise distance for all the solutions (see Table~\ref{tab:results_raw_runs} and~\ref{tab:steer_internal}).

\textbf{Validating \texttt{CodeX-SFT} on a human-curated algorithmic diversity dataset.} \citet{zhang2026rethinking} published 30 hand-curated pairs of Python programs to measure behavioral algorithmic similarity on four types of variations. Type-1 (4 problems) and Type-2 (3 problems) focus on keeping the approach fixed, with variation in either the execution path or the result, respectively. While Type-2 problems do not yield the same results, invalidating them in our algorithmic similarity measurement assumes a prerequisite correctness of the solutions; not all of the four examples in Type-1 can be considered genuinely algorithmically diverse (\textit{e.g.,} changing the order of \texttt{i, j, k} in matrix multiplication). On the other hand, Type-3 (6 problems) and Type-4 (17 problems) align well with our desired algorithmic similarity measures; hence, we place greater emphasis on these two types of problems in this analysis. Type-3 includes two solutions, dissimilar in text, same behavior, and same results (\textit{e.g.,} recursive \textit{vs.} iterative implementations of depth-first search) and Type-4 includes the two solutions that have dissimilar text with the same result but different behavior (\textit{e.g.,} quick sort \textit{vs.} bubble sort). A good measure of algorithmic similarity should assign high scores to Type-3 problems and low scores to Type-4 problems. 

We present the results from~\citet{zhang2026rethinking} as well as the results from our implementation in Table~\ref{tab:behavesim}. While the evaluation dataset is quite small for meaningful statistical analysis, given its hand-curation, the signals it provides still bolster the quality of our trained embedding metric. We found that our trained evaluation metric, \texttt{CodeX-SFT} is able to separate the two types by 0.23 compared to the baseline \texttt{CodeXEmbed} model with 0.08 difference. Since the embedding models were trained on C++ solutions, this is a generalizable result, as the evaluation solutions were in Python, indicating the impact of learning on the contrastive data. The smaller \texttt{Jina} model on the other hand underperforms, which we speculate is a result of overfitting on a small model with 161M parameters compared to the 2.61B parameters in \texttt{CodeXEmbed} model. While their proposed approach \texttt{BehaveSim} remains far ahead with a difference of 0.54, it is expected, as it executes both programs and aligns their intermediate states. This approach limits their similarity measurement to problems with intermediate steps to trace the algorithm's behavior, such as search and clustering, which does not apply to the spectrum of competitive programming problems we investigated in this work.

\input{Tables/behavesim}


%% file: Tables/behavesim.tex
\begin{table}[t]
\centering
\small
\setlength{\tabcolsep}{5pt}
\caption{Measuring algorithmic overlap of our trained evaluation metrics on the hand-curated solution pairs across four types of solution pairs in Python ($n=4,3,6$ and $17$ pairs for each type respectively).}
\label{tab:behavesim}
\begin{tabular}{lccccc}
\toprule
& Type-1 & Type-2 & Type-3 & Type-4 & T3 $-$ T4 \\
\midrule
\multicolumn{6}{l}{\emph{embedding metrics (our implementation)}} \\
\texttt{CodeX-SFT} & 0.93 & 0.85 & 0.87 & 0.65 & +0.23 \\
\texttt{CodeXEmbed} & 0.99 & 0.97 & 0.97 & 0.89 & +0.08 \\
\texttt{jina-SFT} & 0.95 & 0.95 & 0.92 & 0.83 & +0.09 \\
\texttt{jina-embeddings-v2-base-code} & 0.93 & 0.94 & 0.92 & 0.78 & +0.14 \\
\midrule
\multicolumn{6}{l}{\emph{embedding metrics, as reported in~\cite{zhang2026rethinking}}} \\
\texttt{Jina-Code-Embedding} & 0.99 & 0.99 & 0.90 & 0.84 & +0.06 \\
\texttt{Qwen3-Embedding-0.6B} & 0.94 & 0.93 & 0.87 & 0.73 & +0.14 \\
\midrule
\multicolumn{6}{l}{\emph{execution-trajectory metric, as reported in~\cite{zhang2026rethinking}}} \\
\texttt{BehaveSim} & 0.56 & 0.73 & 1.00 & 0.46 & +0.54 \\
\bottomrule
\end{tabular}
\end{table}

%% file: Sections/appx_c_reproducibility.tex
\section{Hyperparameters and Reproducibility Details} \label{appx:hyperparams}

Except for the temperature variation study, all the models were sampled at $T{=}0.8$, $\mathrm{top}_p{=}0.95$, $N{=}50$ samples per problem over the $1{,}000$ test split problems. The maximum token limit varied across models to ensure less than 10\% truncation (see Table~\ref{tab:results_raw_runs} for individual model hyperparameters). Every solution was compiled with \texttt{g++ -O2 -std=c++17} and run against the 15 test cases provided by the original \texttt{CodeContests+}~\cite{} benchmark. Program dependence graphs were extracted with the Joern library\footnote{Joern library URL: \url{https://docs.joern.io/export/}. We removed some stylistic attributes from the PDG graph post hoc to make it less syntax-dependent. Each PDG node was relabeled by dropping the line numbers, replacing user-defined variables with `VAR' and numeric literals by `NUM'; and the variable name was dropped from the edge dependence to only contain the dependence type (control or data) and the direction.} The similarity between graphs was computed using a three-iteration Weisfeiler-Lehman subtree kernel~\cite{shervashidze2011weisfeiler}. The trained similarity metric \texttt{CodeX-SFT} was fine-tuned with a maximum of 2,048 input tokens. All local experiments were run on four NVIDIA A100 40GB GPUs.

%% file: Sections/appx_d_results.tex
\section{Results (Extended)} \label{appx:results}

We provide the extended results for the experiments described in Section~\ref{sec:rq1} and Section~\ref{sec:rq2}. Table~\ref{tab:results_raw_runs} reports the uncontrolled zero-shot generation results for the 24 LLMs with a human reference. Table~\ref{tab:steer_internal} reports the controlled diverse code generation results for the three open-weight LLMs for the three strategies: temperature variation, verbalized sampling, and iterative constrained generation. 

\begin{figure}[h]
    \centering
    \includegraphics[width=0.85\linewidth]{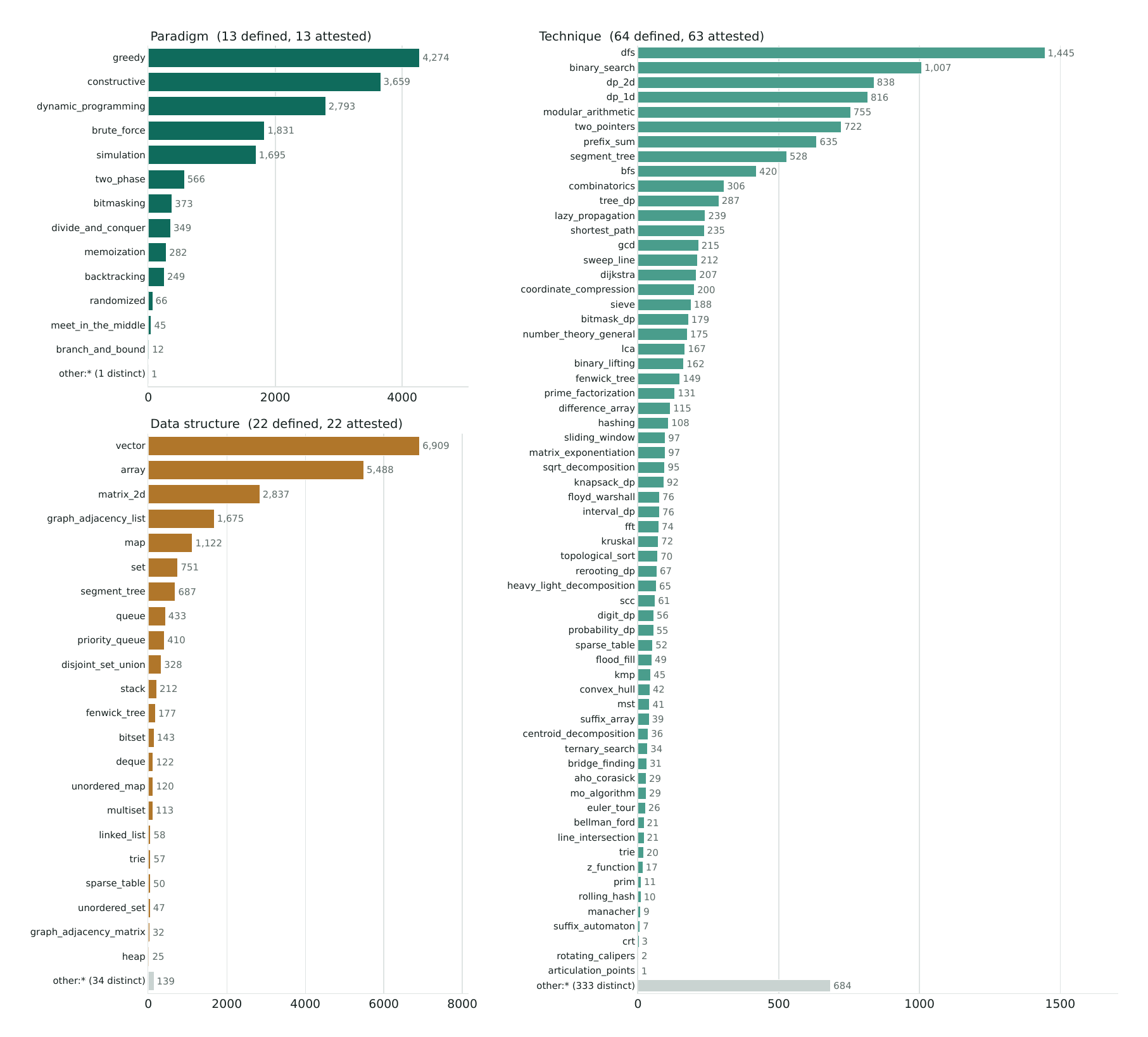}
    \caption{Distribution of ontology labels in the two C++ solutions per programming problem in the algorithmic diversity dataset.}
    \label{fig:ontology-distribution}
\end{figure}

\input{Tables/main_results_raw}

\input{Tables/result_steer_internal}

\newpage

\begin{figure}
    \centering
    \includegraphics[width=\linewidth]{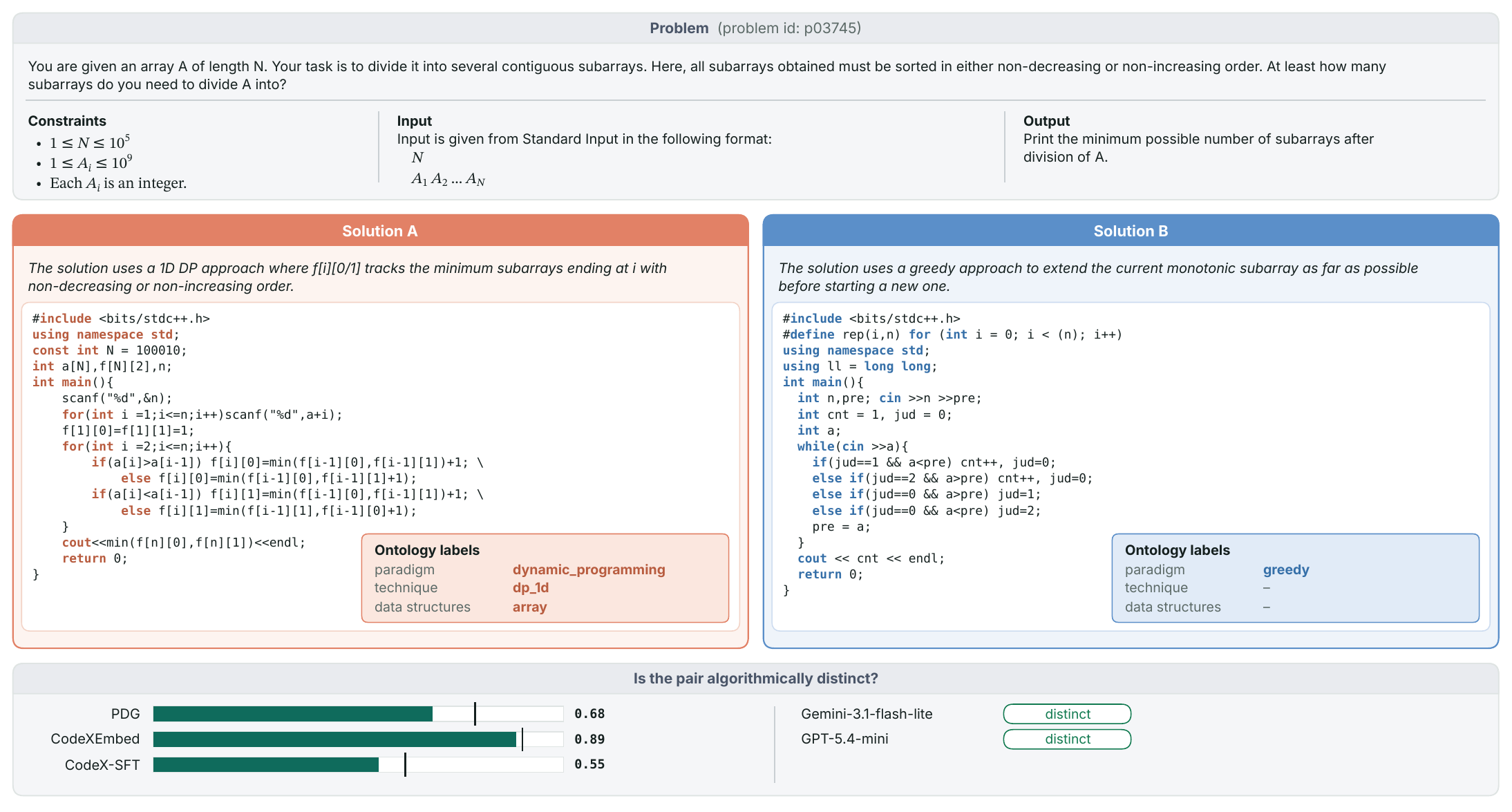}
    \caption{Illustrative example of \textit{\textbf{strong}} algorithmic diversity from the train split. The two solutions have different programming paradigms. The bar graphs on the bottom left provide the code similarity of different similarity metrics, with the ticks indicating a threshold calibrated on the 10th percentile of randomly selected 150 train instances. Similarity below the threshold is considered algorithmically diverse as per the metric. We also provide the pairwise LLM judgments from the second filtration stage on the bottom right.}
    \label{fig:example-strong}
\end{figure}

\begin{figure}
    \centering
    \includegraphics[width=\linewidth]{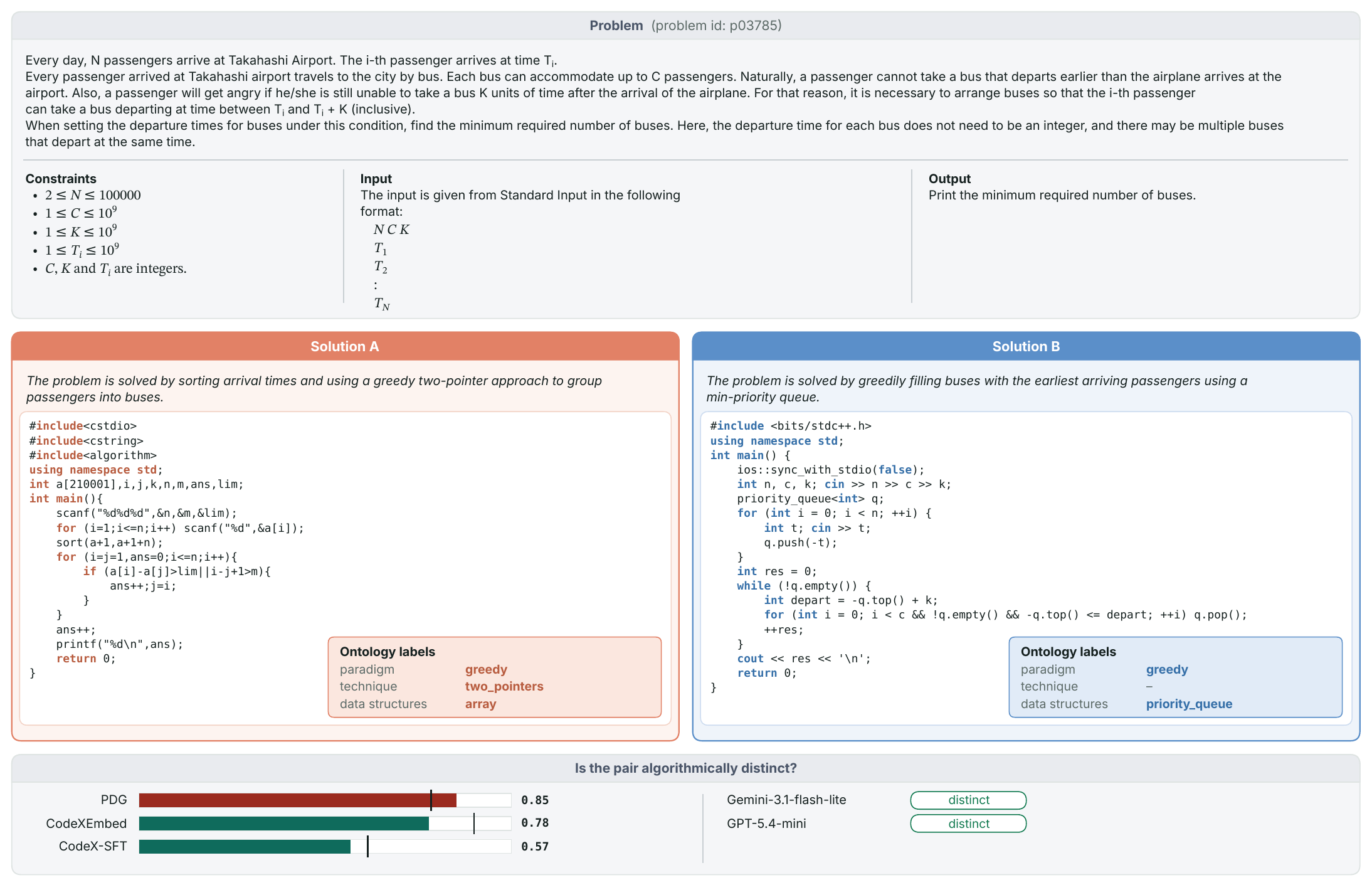}
    \caption{Illustrative example of \textit{\textbf{medium}} algorithmic diversity from the train split. The two solutions have the same programming paradigm, but different techniques. The bar graphs at the bottom left show code similarity across different similarity metrics, with the ticks indicating a threshold calibrated to the 10th percentile of randomly selected 150 training instances. Similarity below the threshold is considered algorithmically diverse as per the metric. We also provide the pairwise LLM judgments from the second filtration stage on the bottom right.}
    \label{fig:example-medium}
\end{figure}

\begin{figure}
    \centering
    \includegraphics[width=\linewidth]{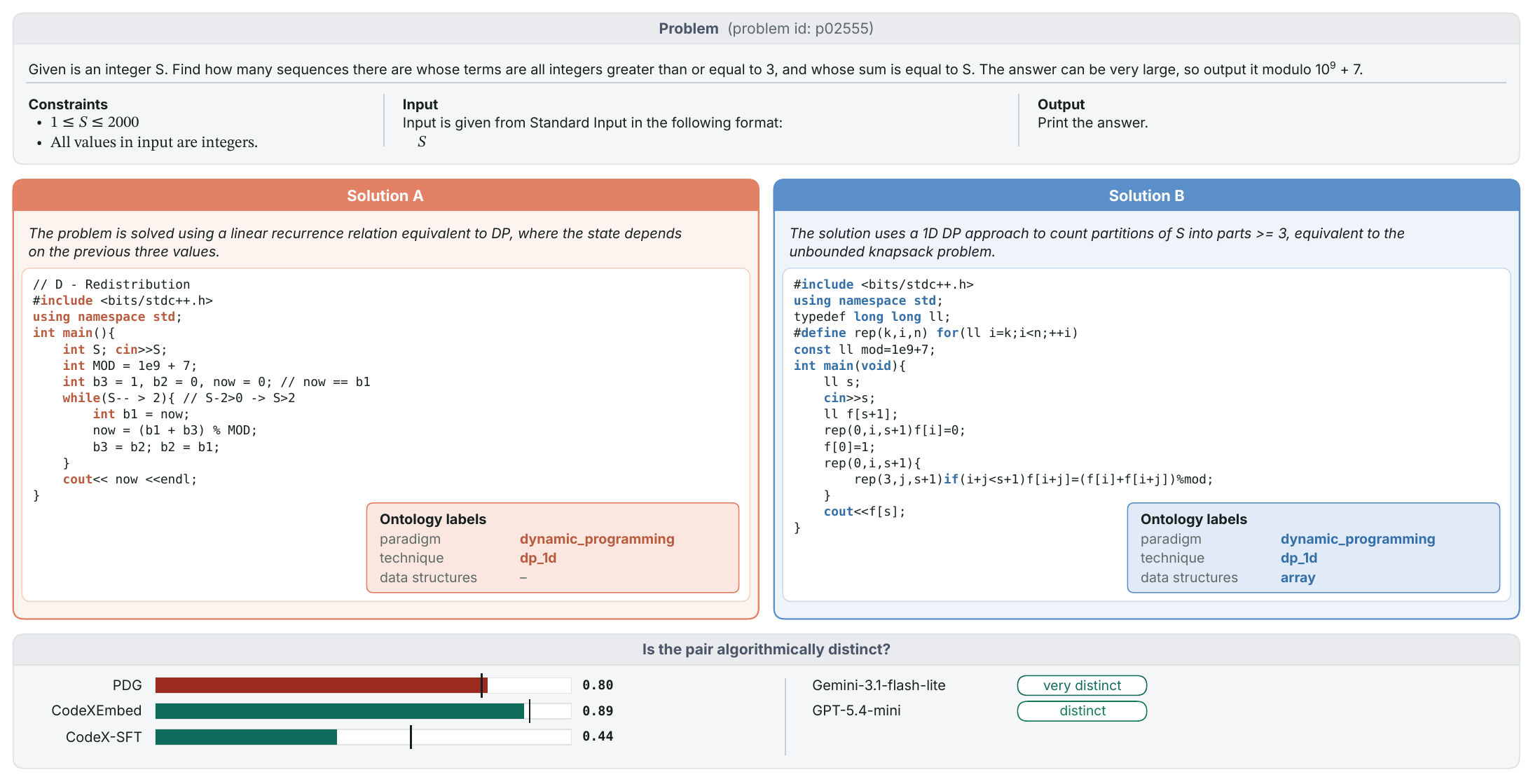}
    \caption{Illustrative example of \textit{\textbf{weak}} algorithmic diversity from the train split. The two solutions use the same programming paradigm and techniques but different data structures. The bar graphs at the bottom left show code similarity across different similarity metrics, with the ticks indicating a threshold calibrated to the 10th percentile of randomly selected 150 training instances. Similarity below the threshold is considered algorithmically diverse as per the metric. We also provide the pairwise LLM judgments from the second filtration stage on the bottom right.}
    \label{fig:example-weak}
\end{figure}

\begin{figure}
    \centering
    \includegraphics[width=\linewidth]{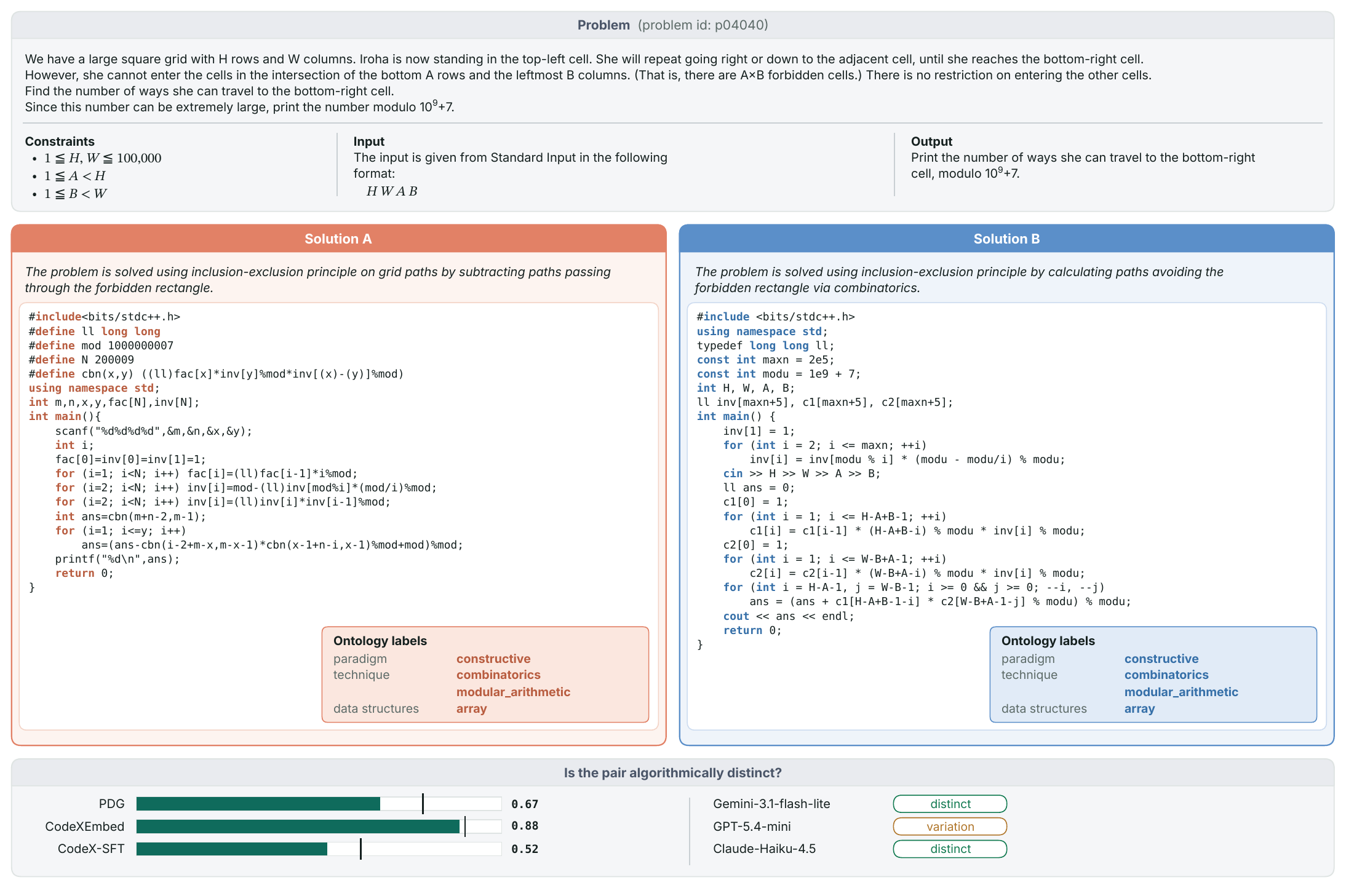}
    \caption{Illustrative example of \textit{\textbf{identical}} algorithmic diversity from the train split. The two solutions share the same programming paradigm, techniques, and data structures, and pass through all the filtration steps. The bar graphs at the bottom left show code similarity across different similarity metrics, with the ticks indicating a threshold calibrated to the 10th percentile of randomly selected 150 training instances. Similarity below the threshold is considered algorithmically diverse as per the metric. We also provide the pairwise LLM judgments from the second filtration stage on the bottom right.}
    \label{fig:example-identical}
\end{figure}

%% file: Tables/main_results_raw.tex
\begin{table}[t]
\centering
\small
\setlength{\tabcolsep}{4pt}
\caption{Algorithmic diversity results for unconditioned zero-shot generations on the 1,000 test set programming problems. We present three success statistics: \textit{pass rate} - percentage of generated solutions that pass all test cases, \textit{Pass@N} - number of instances with at least one correct generation, and \textit{Pass$_{\geq2}$@N} - number of instances with at least two correct solutions to compute within problem diversity. We also provide the generation statistics, including the median (\textit{med}) and maximum (\textit{max}) number of output tokens, as well as the percentage of solutions that were truncated (\textit{trunc}). For per-problem population diversity, we present the self-diversity (\textbf{self-div}) of generated solutions as the average distance of all correct solutions using two distance measurements: program dependence graphs (\textit{PDG}) and the trained embedding model \textit{CodeX-SFT} (Section~\ref{appx:trained-embedding}). We also provide the $\mathrm{Vendi}_{\texttt{CodeX-SFT}}$ (Eq.~\ref{eq:vendi}) scores on two settings, \textit{matched} and \textit{all solutions}. \textit{Matched} $\mathrm{Vendi}$ score subsamples $k^*=\text{min}(k_{\text{LLM}},k_{\text{hum}})$ number of solutions from the LLM-generated solutions and the human references over 200 draws without replacement, whereas the \textit{all solutions} $\mathrm{Vendi}$ score is computed at the entire pool size of each solution population. We also provide the human reference diversity scores for all 1,000 test instances.}
\label{tab:results_raw_runs}
\resizebox{\linewidth}{!}{%
\begin{tabular}{lccccccccccc}
\toprule
& \multicolumn{3}{c}{\textbf{Success Statistics}} & \multicolumn{3}{c}{\textbf{\#tokens}} & \multicolumn{2}{c}{\textbf{self-div}}
& \multicolumn{3}{c}{$\mathrm{Vendi}_{\texttt{CodeX-SFT}}$\textnormal{($k^*$ matched / all solutions)}} \\
\cmidrule(lr){2-4}\cmidrule(lr){5-7}\cmidrule(lr){8-9}\cmidrule(lr){10-12}
Model & \textit{Pass rate} & \textit{Pass@N} & \textit{Pass$_{\geq2}$@N} & \textit{med} & \textit{max} & \textit{trunc (\%)} &
\textit{PDG} & \textit{CodeX-SFT} & $k^{*} / k_{\mathrm{LLM}}$ & $\mathrm{VS}^{\mathrm{LLM}}$ & $\mathrm{VS}^{\mathrm{hum}}$ \\
\midrule
\texttt{Gemini-3.8-flash} & 80.9 & 954 & 945 & -- & -- & 0.7 & 0.046 & 0.175 & 24.7 / 42.8 & 4.18 / 4.84 & 13.29 / 14.25 \\
\texttt{Gemma-4-31b} & 76.8 & 927 & 913 & 1,073 & 8,192 & 5.1 & 0.049 & 0.173 & 24.7 / 42.0 & 4.14 / 4.75 & 13.15 / 14.23 \\
\texttt{Gemma-4-26B} & 74.0 & 929 & 917 & 2,447 & 16,384 & 7.2 & 0.058 & 0.171 & 24.3 / 40.3 & 3.50 / 3.96 & 13.01 / 14.26 \\
\texttt{Gemma-4-12b} & 64.1 & 881 & 858 & 1,576 & 16,384 & 10.0 & 0.054 & 0.178 & 23.3 / 37.3 & 3.71 / 4.13 & 12.24 / 14.16 \\
\texttt{Gemini-flash-lite} & 53.6 & 817 & 782 & -- & -- & 0.0 & 0.050 & 0.187 & 21.9 / 34.2 & 4.32 / 4.75 & 11.29 / 13.80 \\
\texttt{Gemma-4-E4B} & 44.7 & 736 & 692 & 2,132 & 8,192 & 0.8 & 0.066 & 0.195 & 21.6 / 32.2 & 4.23 / 4.64 & 10.86 / 13.57 \\
\texttt{Phi-4} & 31.4 & 678 & 618 & 356 & 2,048 & 0.1 & 0.066 & 0.169 & 18.5 / 25.3 & 3.22 / 3.33 & 8.79 / 13.18 \\
\texttt{Qwen3-Coder-30B} & 25.8 & 516 & 463 & 824 & 8,192 & 5.1 & 0.067 & 0.160 & 19.5 / 27.8 & 3.22 / 3.34 & 9.37 / 13.05 \\
\texttt{Qwen2.5-Coder-32B} & 19.2 & 409 & 366 & 271 & 2,048 & 0.1 & 0.058 & 0.158 & 18.4 / 26.2 & 2.97 / 3.07 & 8.47 / 12.52 \\
\texttt{DeepSeek-V2-Lite} & 14.6 & 356 & 321 & 392 & 2,048 & 0.1 & 0.051 & 0.129 & 16.9 / 22.7 & 2.38 / 2.43 & 7.74 / 12.09 \\
\texttt{Codestral-22B} & 11.9 & 388 & 327 & 416 & 2,048 & 0.2 & 0.086 & 0.207 & 14.4 / 18.0 & 3.46 / 3.62 & 6.79 / 12.08 \\
\texttt{Qwen2.5-7B} & 8.8 & 308 & 246 & 274 & 2,048 & 0.1 & 0.073 & 0.169 & 14.6 / 17.6 & 2.93 / 2.99 & 6.31 / 11.73 \\
\texttt{Qwen2.5-Coder-7B} & 8.7 & 271 & 224 & 271 & 2,048 & 0.2 & 0.071 & 0.170 & 15.6 / 19.3 & 3.05 / 3.14 & 7.01 / 11.80 \\
\texttt{Starcoder2-15b} & 7.3 & 270 & 218 & 351 & 2,048 & 0.2 & 0.097 & 0.195 & 13.7 / 16.4 & 3.24 / 3.34 & 6.26 / 11.75 \\
\texttt{Mistral-Nemo} & 5.5 & 244 & 186 & 257 & 2,048 & 0.0 & 0.080 & 0.186 & 12.7 / 14.6 & 3.07 / 3.14 & 5.80 / 11.47 \\
\texttt{Codegemma-7b-it} & 5.2 & 196 & 152 & 236 & 2,048 & 0.0 & 0.074 & 0.169 & 14.1 / 16.7 & 2.95 / 3.01 & 6.26 / 11.35 \\
\texttt{Llama-3.1-8B} & 4.7 & 221 & 175 & 338 & 2,048 & 0.4 & 0.104 & 0.230 & 11.5 / 13.3 & 3.54 / 3.60 & 5.20 / 10.81 \\
\texttt{CodeLlama-34B} & 3.5 & 171 & 137 & 360 & 2,048 & 0.4 & 0.095 & 0.202 & 10.9 / 12.5 & 3.13 / 3.23 & 5.07 / 10.70 \\
\texttt{CodeQwen1.5-7B} & 3.2 & 186 & 139 & 403 & 2,048 & 1.6 & 0.114 & 0.245 & 9.8 / 11.0 & 3.33 / 3.41 & 4.49 / 11.17 \\
\texttt{CodeLlama-13B} & 3.1 & 148 & 117 & 303 & 2,048 & 0.2 & 0.092 & 0.207 & 11.4 / 12.9 & 3.08 / 3.11 & 5.24 / 10.92 \\
\texttt{Llama-3.2-3B} & 2.7 & 158 & 116 & 305 & 2,048 & 0.3 & 0.097 & 0.227 & 10.6 / 11.3 & 3.26 / 3.27 & 4.63 / 10.36 \\
\texttt{CodeLlama-7B} & 2.5 & 130 & 101 & 344 & 2,048 & 0.3 & 0.088 & 0.206 & 10.3 / 12.3 & 2.98 / 3.02 & 4.77 / 10.47 \\
\texttt{Mistral-7B-v0.3} & 1.7 & 118 & 77 & 455 & 2,048 & 0.1 & 0.086 & 0.222 & 9.6 / 10.8 & 3.07 / 3.09 & 4.20 / 9.77 \\
\texttt{Qwen2.5-0.5B} & 0.9 & 79 & 49 & 247 & 2,048 & 1.5 & 0.107 & 0.257 & 7.8 / 8.4 & 3.04 / 3.04 & 3.59 / 9.54 \\
\midrule
\texttt{Human reference} & -- & 1,000 & 1,000 & -- & -- & -- & 0.215 & 0.374 & -- / 26.6 & -- & -- / 14.27 \\
\bottomrule
\end{tabular}
}
\end{table}

%% file: Tables/result_steer_internal.tex
\begin{table}[t]
\centering
\small
\setlength{\tabcolsep}{4pt}
\caption{Algorithmic diversity of diversity conditioned generation approaches on the 1,000 test set programming problems. We present the results for three open-weight models over three strategies: temperature variation with 50 generations (\textit{i.i.d.} denotes the standard T=0.8 setting), \textit{verbalized} sampling (10 independent calls per problem asking to generate five solutions per call), and \textit{iterative} generation (10 independent chains per problem with five iterative constrained generations per generation chain). All conditions are evaluated on the same 50 generated solutions, and the statistics and diversity measurements are the same as described in Table~\ref{tab:results_raw_runs}.}
\label{tab:steer_internal}
\resizebox{\linewidth}{!}{%
\begin{tabular}{llccccccccccc}
\toprule
& & \multicolumn{3}{c}{\textbf{Success Statistics}} & \multicolumn{3}{c}{\textbf{\#tokens}}
& \multicolumn{2}{c}{\textbf{self-div}} & \multicolumn{3}{c}{$\mathrm{Vendi}_{\texttt{CodeX-SFT}}$\textnormal{($k^*$ matched / all solutions)}} \\
\cmidrule(lr){3-5}\cmidrule(lr){6-8}\cmidrule(lr){9-10}\cmidrule(lr){11-13}
Model & Sampling & \textit{Pass rate} & \textit{Pass$_{\geq2}$@N} & \textit{k} &
\textit{med} & \textit{max} & \textit{trunc (\%)} &
\textit{PDG} & \textit{CodeX-SFT} & $k^{*} / k_{\mathrm{LLM}}$ & $\mathrm{VS}^{\mathrm{LLM}}$ & $\mathrm{VS}^{\mathrm{hum}}$ \\
\midrule
\texttt{Phi-4} & i.i.d. & 31.4 & 618 & 25.3 & 356 & 2,048 & 0.09 & 0.066 & 0.169 & 18.5 / 25.3 & 3.22 / 3.33 & 8.79 / 13.18 \\
 & temp 0.2 & 33.5 & 538 & 31.1 & 347 & 2,048 & 0.08 & 0.032 & 0.087 & 20.6 / 31.1 & 1.85 / 1.89 & 9.56 / 12.95 \\
 & temp 1.2 & 27.6 & 587 & 23.4 & 380 & 2,048 & 1.76 & 0.079 & 0.197 & 17.7 / 23.4 & 3.73 / 3.85 & 8.23 / 12.96 \\
 & iterative & 24.0 & 604 & 19.8 & 370 & 2,048 & 0.02 & 0.072 & 0.219 & 16.7 / 19.8 & 4.16 / 4.31 & 8.01 / 13.06 \\
 & verbalized & 22.7 & 561 & 20.2 & 1,807 & 13,000 & 0.01 & 0.067 & 0.191 & 16.8 / 20.2 & 3.60 / 3.71 & 8.07 / 13.00 \\
\midrule
\texttt{Gemma-4-E4B} & i.i.d. & 44.7 & 692 & 32.2 & 2,132 & 8,192 & 0.81 & 0.066 & 0.195 & 21.6 / 32.2 & 4.23 / 4.64 & 10.86 / 13.57 \\
 & temp 0.2 & 44.9 & 681 & 33.0 & 2,016 & 8,192 & 0.69 & 0.052 & 0.172 & 21.7 / 33.0 & 3.46 / 3.73 & 10.91 / 13.61 \\
 & temp 1.2 & 43.3 & 693 & 31.2 & 2,377 & 8,192 & 1.29 & 0.078 & 0.211 & 21.4 / 31.2 & 4.71 / 5.18 & 10.72 / 13.62 \\
 & iterative & 44.5 & 707 & 31.4 & 1,760 & 8,192 & 0.20 & 0.071 & 0.252 & 21.9 / 31.4 & 6.18 / 7.00 & 11.09 / 13.65 \\
 & verbalized & 38.4 & 672 & 17.7 & 6,242 & 16,384 & 2.86 & 0.090 & 0.275 & 15.8 / 17.7 & 5.79 / 5.97 & 8.41 / 13.55 \\
\midrule
\texttt{Gemma-4-31b} & i.i.d. & 76.8 & 913 & 42.0 & 1,073 & 8,192 & 5.09 & 0.049 & 0.173 & 24.7 / 42.0 & 4.14 / 4.75 & 13.15 / 14.23 \\
 & temp 0.2 & 76.9 & 909 & 42.3 & 1,073 & 8,192 & 5.59 & 0.040 & 0.150 & 24.7 / 42.3 & 3.42 / 3.83 & 13.18 / 14.22 \\
 & temp 1.2 & 76.4 & 918 & 41.6 & 1,077 & 8,192 & 4.93 & 0.054 & 0.186 & 24.7 / 41.6 & 4.58 / 5.31 & 13.16 / 14.25 \\
 & iterative & 74.2 & 920 & 40.3 & 1,194 & 8,192 & 1.76 & 0.083 & 0.308 & 24.9 / 40.3 & 8.83 / 10.97 & 13.31 / 14.22 \\
 & verbalized & 73.0 & 907 & 39.2 & 3,546 & 16,384 & 0.17 & 0.077 & 0.280 & 24.4 / 39.2 & 7.45 / 8.88 & 13.01 / 14.22 \\
\bottomrule
\end{tabular}}
\end{table}

%% file: Sections/appx_z_prompts.tex
\section{Prompts} \label{appx:prompts}

\begin{figure}
    \centering
    \includegraphics[width=\linewidth]{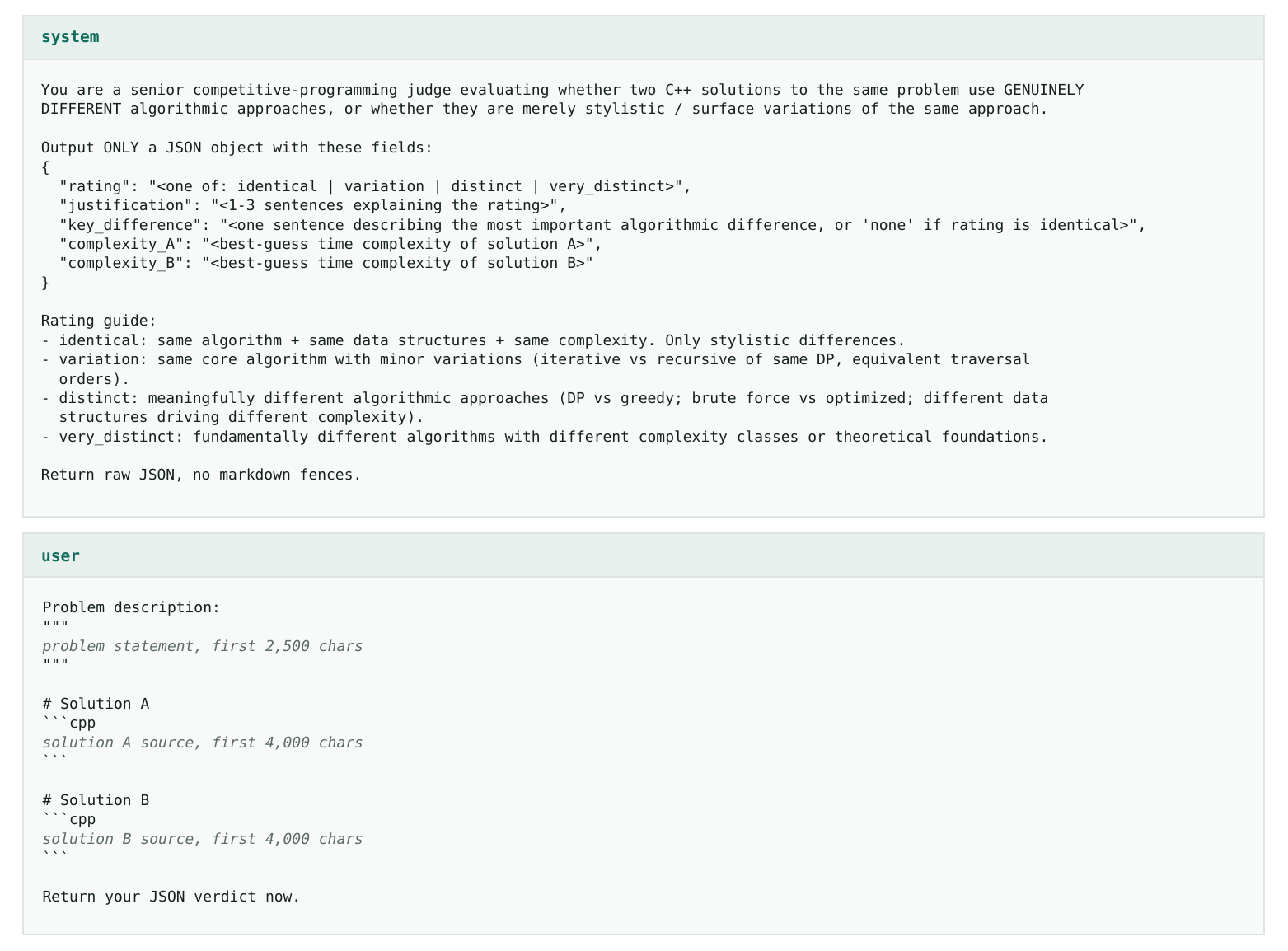}
    \caption{The LLM-judge prompt used to obtain the 4-way ordinal classification of algorithmic similarity between two programs for the same problem.}
    \label{fig:prompt-judge}
\end{figure}

\begin{figure}
    \centering
    \includegraphics[width=\linewidth]{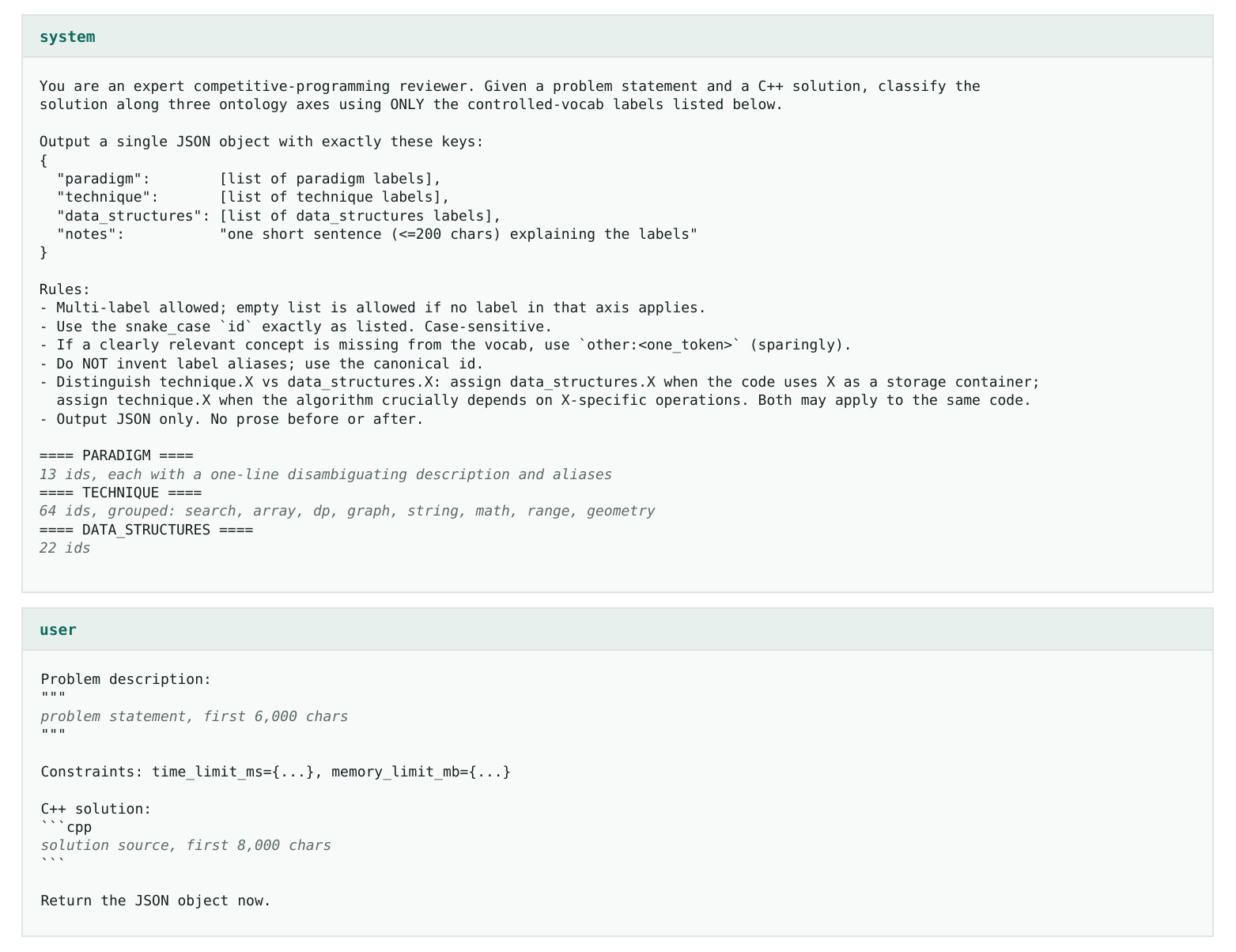}
    \caption{The prompt used to obtain the ontology labels for the two representative solutions in the final dataset. }
    \label{fig:prompt-ontology}
\end{figure}

\begin{figure}
    \centering
    \includegraphics[width=\linewidth]{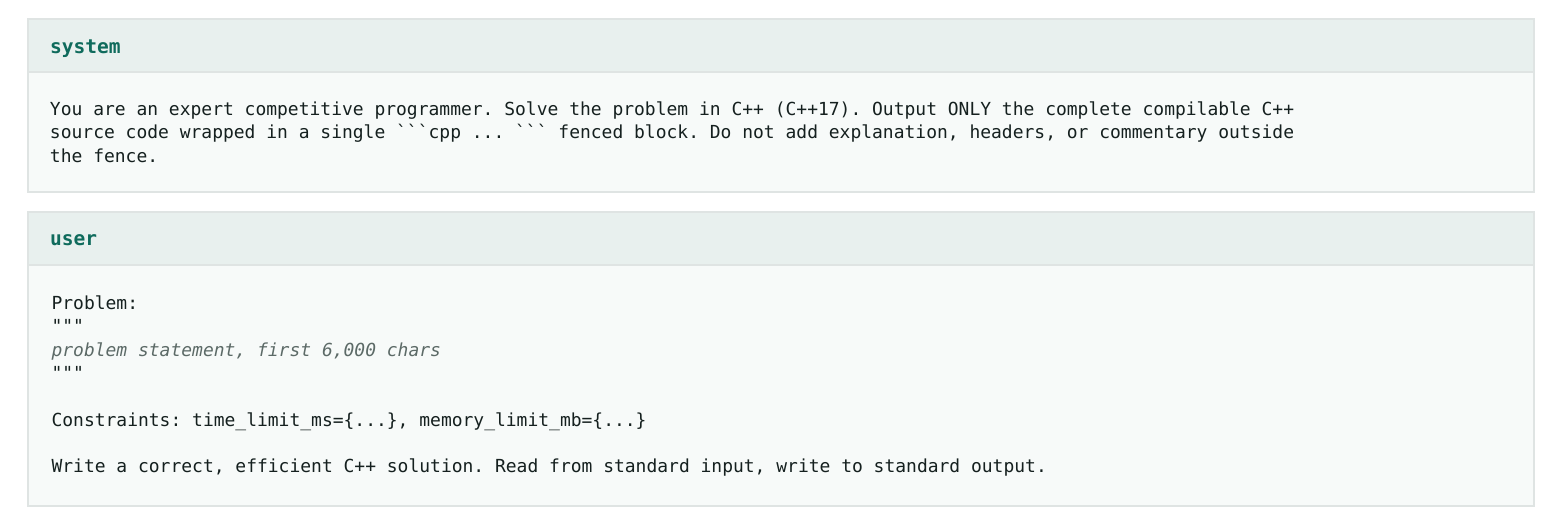}
    \caption{The prompt used to generate \textit{i.i.d.} zero-shot solutions for different temperature values.}
    \label{fig:prompt-iid}
\end{figure}

\begin{figure}
    \centering
    \includegraphics[width=\linewidth]{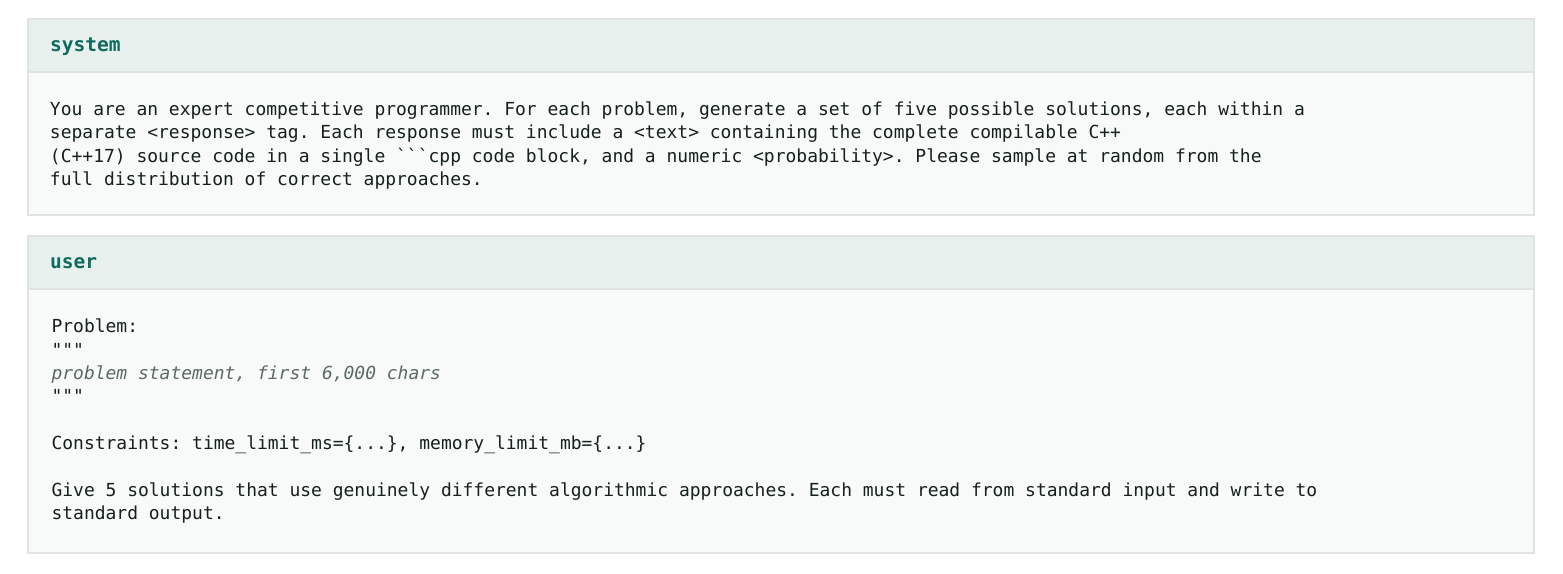}
    \caption{The prompt used to generate the solutions using the verbalized sampling approach~\cite{zhang2025verbalized}. We limit to 5 outputs per generation and re-run it 10 times to obtain 50 solutions for direct comparison with the \textit{i.i.d.} setting.}
    \label{fig:prompt-verbalized}
\end{figure}

\begin{figure}
    \centering
    \includegraphics[width=\linewidth]{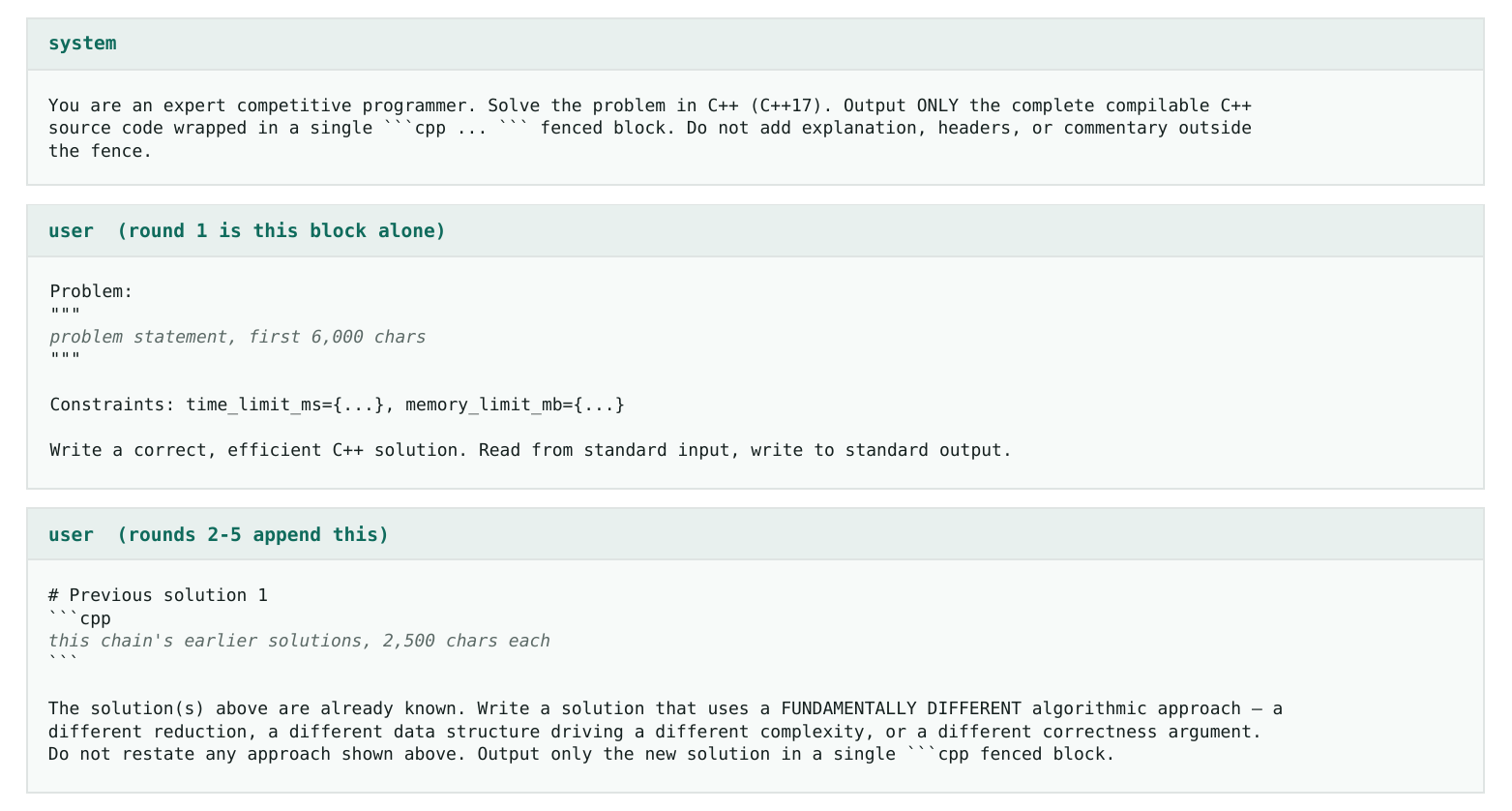}
    \caption{The prompt used to iteratively generate diverse solutions. We limit the iterations to 5 generations and re-run it 10 times to obtain 50 solutions for direct comparison with the \textit{i.i.d.} setting.}
    \label{fig:prompt-iterative}
\end{figure}

\begin{figure}
    \centering
    \includegraphics[width=\linewidth]{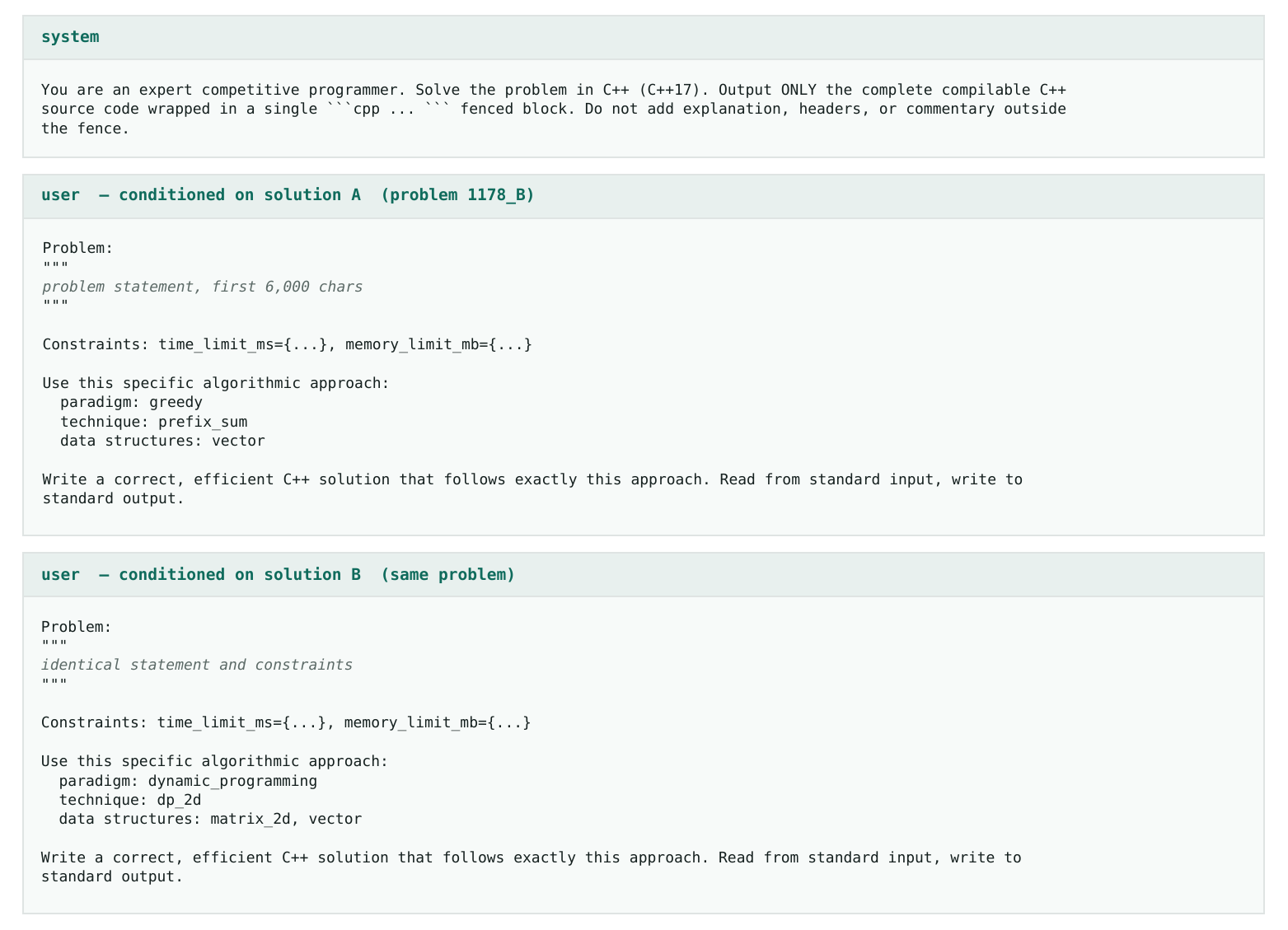}
    \caption{The prompts used to generate the steer (A+B) solutions that are instructed to adhere to a particular algorithmic strategy.}
    \label{fig:prompt-steer-ab}
\end{figure}

We provide the prompts we have used throughout this work in this Section. Figure~\ref{fig:prompt-ontology} illustrates the prompt used to classify the solutions' used algorithms following the proposed ontology. For evaluation, Figure~\ref{fig:prompt-judge} presents the four-way ordinal evaluation prompt. For our generation, Figures~\ref{fig:prompt-iid}, ~\ref{fig:prompt-verbalized} and ~\ref{fig:prompt-iterative} present the \textit{i.i.d.}, \textit{iterative} and \textit{verbalized} prompts used to generate solutions. Figure~\ref{fig:prompt-steer-ab} shows the prompt used to obtain the steer (A+B) solutions following specific solution ontology.

%% file: main.bib
@article{liu2024codexembed,
  title={Codexembed: A generalist embedding model family for multiligual and multi-task code retrieval},
  author={Liu, Ye and Meng, Rui and Joty, Shafiq and Savarese, Silvio and Xiong, Caiming and Zhou, Yingbo and Yavuz, Semih},
  journal={arXiv preprint arXiv:2411.12644},
  year={2024}
}

@article{friedman2022vendi,
  title={The vendi score: A diversity evaluation metric for machine learning},
  author={Friedman, Dan and Dieng, Adji Bousso},
  journal={arXiv preprint arXiv:2210.02410},
  year={2022}
}

@inproceedings{li2025coir,
  title={Coir: A comprehensive benchmark for code information retrieval models},
  author={Li, Xiangyang and Dong, Kuicai and Lee, Yi Quan and Xia, Wei and Zhang, Hao and Dai, Xinyi and Wang, Yasheng and Tang, Ruiming},
  booktitle={Proceedings of the 63rd Annual Meeting of the Association for Computational Linguistics (Volume 1: Long Papers)},
  pages={22074--22091},
  year={2025}
}

@inproceedings{lee2025diversely,
  title={How Diversely Can Language Models Solve Problems? Exploring the Algorithmic Diversity of Model-Generated Code.},
  author={Lee, Seonghyeon and Chon, Heejae and Jang, Joonwon and Lee, Dongha and Yu, Hwanjo},
  booktitle={EMNLP (Findings)},
  pages={152--167},
  year={2025}
}

@article{zhang2025verbalized,
  title={Verbalized sampling: How to mitigate mode collapse and unlock llm diversity},
  author={Zhang, Jiayi and Yu, Simon and Chong, Derek and Sicilia, Anthony and Tomz, Michael R and Manning, Christopher D and Shi, Weiyan},
  journal={arXiv preprint arXiv:2510.01171},
  year={2025}
}

@inproceedings{
zhang2026rethinking,
title={Rethinking Code Similarity for Automated Algorithm Design with {LLM}s},
author={Rui Zhang and Zhichao Lu},
booktitle={The Fourteenth International Conference on Learning Representations},
year={2026},
url={https://openreview.net/forum?id=HIUqeO9OOr}
}

@article{shervashidze2011weisfeiler,
  title={Weisfeiler-lehman graph kernels.},
  author={Shervashidze, Nino and Schweitzer, Pascal and Van Leeuwen, Erik Jan and Mehlhorn, Kurt and Borgwardt, Karsten M},
  journal={Journal of Machine Learning Research},
  volume={12},
  number={9},
  year={2011}
}

@article{ren2020codebleu,
  title={Codebleu: a method for automatic evaluation of code synthesis},
  author={Ren, Shuo and Guo, Daya and Lu, Shuai and Zhou, Long and Liu, Shujie and Tang, Duyu and Sundaresan, Neel and Zhou, Ming and Blanco, Ambrosio and Ma, Shuai},
  journal={arXiv preprint arXiv:2009.10297},
  year={2020}
}

@inproceedings{feng2020codebert,
  title={Codebert: A pre-trained model for programming and natural languages},
  author={Feng, Zhangyin and Guo, Daya and Tang, Duyu and Duan, Nan and Feng, Xiaocheng and Gong, Ming and Shou, Linjun and Qin, Bing and Liu, Ting and Jiang, Daxin and others},
  booktitle={Findings of the association for computational linguistics: EMNLP 2020},
  pages={1536--1547},
  year={2020}
}

@inproceedings{shah2025evolution,
  title={Evolution of Programmers' Trust in Generative AI Programming Assistants},
  author={Shah, Anshul and Rexin, Thomas and Tomson, Elena and Griswold, William G and Porter, Leo and Soosai Raj, Adalbert Gerald},
  booktitle={Proceedings of the 25th Koli Calling International Conference on Computing Education Research},
  pages={1--11},
  year={2025}
}

@article{khati2025mapping,
  title={Mapping the trust terrain: LLMs in Software Engineering-insights and perspectives},
  author={Khati, Dipin and Liu, Yijin and Palacio, David N and Zhang, Yixuan and Poshyvanyk, Denys},
  journal={ACM Transactions on Software Engineering and Methodology},
  year={2025},
  publisher={ACM New York, NY}
}

@article{shao2024deepseekmath,
  title={Deepseekmath: Pushing the limits of mathematical reasoning in open language models},
  author={Shao, Zhihong and Wang, Peiyi and Zhu, Qihao and Xu, Runxin and Song, Junxiao and Bi, Xiao and Zhang, Haowei and Zhang, Mingchuan and Li, YK and Wu, Yang and others},
  journal={arXiv preprint arXiv:2402.03300},
  year={2024}
}

@article{lambert2024tulu,
  title={Tulu 3: Pushing frontiers in open language model post-training},
  author={Lambert, Nathan and Morrison, Jacob and Pyatkin, Valentina and Huang, Shengyi and Ivison, Hamish and Brahman, Faeze and Miranda, Lester James V and Liu, Alisa and Dziri, Nouha and Lyu, Shane and others},
  journal={arXiv preprint arXiv:2411.15124},
  year={2024}
}

@article{wang2026code,
  title={Code-A1: Adversarial evolving of code LLM and test LLM via reinforcement learning},
  author={Wang, Aozhe and Yan, Yuchen and Zhou, Nan and Lu, Zhengxi and Lu, Weiming and Xiao, Jun and Zhuang, Yueting and Shen, Yongliang},
  journal={arXiv preprint arXiv:2603.15611},
  year={2026}
}

@article{li2026exploring,
  title={Exploring Pass-Rate Reward in Reinforcement Learning for Code Generation},
  author={Li, Xin-Ye and Liu, Ren-Biao and Zhang, Yun-Ji and Sun, Hui and Xie, Zheng and Li, Ming},
  journal={arXiv preprint arXiv:2605.02944},
  year={2026}
}

@inproceedings{zeng2025acecoder,
  title={Acecoder: Acing coder rl via automated test-case synthesis},
  author={Zeng, Huaye and Jiang, Dongfu and Wang, Haozhe and Nie, Ping and Chen, Xiaotong and Chen, Wenhu},
  booktitle={Proceedings of the 63rd Annual Meeting of the Association for Computational Linguistics (Volume 1: Long Papers)},
  pages={12023--12040},
  year={2025}
}

@inproceedings{
narula2026the,
title={The Invisible Lottery: How Subtle Cues Steer Algorithm Choice in {LLM} Code Generation},
author={Akanksha Narula and Mofasshara Binte Rafique and Laurent Bindschaedler},
booktitle={Forty-third International Conference on Machine Learning},
year={2026},
url={https://openreview.net/forum?id=JrVfxa6aLL}
}

@inproceedings{ye2025assessing,
  title={Assessing the creativity of llms in proposing novel solutions to mathematical problems},
  author={Ye, Junyi and Gu, Jingyi and Zhao, Xinyun and Yin, Wenpeng and Wang, Guiling},
  booktitle={Proceedings of the AAAI Conference on Artificial Intelligence},
  volume={39},
  number={24},
  pages={25687--25696},
  year={2025}
}

@inproceedings{lu2025benchmarking,
  title={Benchmarking language model creativity: A case study on code generation},
  author={Lu, Yining and Wang, Dixuan and Li, Tianjian and Jiang, Dongwei and Khudanpur, Sanjeev and Jiang, Meng and Khashabi, Daniel},
  booktitle={Proceedings of the 2025 Conference of the Nations of the Americas Chapter of the Association for Computational Linguistics: Human Language Technologies (Volume 1: Long Papers)},
  pages={2776--2794},
  year={2025}
}

@article{almeida2026effectiveness,
  title={Effectiveness of LLM-based Software Diversity for Reliability Improvement--an Empirical Study},
  author={Almeida, Gabriel and Gashi, Ilir and Stankovic, Vladimir and Campos, Jo{\~a}o R},
  journal={arXiv preprint arXiv:2607.03174},
  year={2026}
}

@article{florian2026beyond,
  title={Beyond pass@ k: Redundancy-Aware RLVR for Multi-Sample Code Generation},
  author={Le Bronnec, Florian and Verine, Alexandre and Yokota, Rio and Negrevergne, Benjamin},
  journal={arXiv preprint arXiv:2605.28022},
  year={2026}
}

@article{jiang2026survey,
  title={A survey on large language models for code generation},
  author={Jiang, Juyong and Wang, Fan and Shen, Jiasi and Kim, Sungju and Kim, Sunghun},
  journal={ACM Transactions on Software Engineering and Methodology},
  volume={35},
  number={2},
  pages={1--72},
  year={2026},
  publisher={ACM New York, NY}
}

@article{dehaerne2022code,
  title={Code generation using machine learning: A systematic review},
  author={Dehaerne, Enrique and Dey, Bappaditya and Halder, Sandip and De Gendt, Stefan and Meert, Wannes},
  journal={Ieee Access},
  volume={10},
  pages={82434--82455},
  year={2022},
  publisher={IEEE}
}

@article{chen2024survey,
  title={A survey on evaluating large language models in code generation tasks},
  author={Chen, Liguo and Guo, Qi and Jia, Hongrui and Zeng, Zhengran and Wang, Xin and Xu, Yijiang and Wu, Jian and Wang, Yidong and Gao, Qing and Wang, Jindong and others},
  journal={arXiv preprint arXiv:2408.16498},
  year={2024}
}

@book{mccarthy1964formal,
  title={A formal description of a subset of ALGOL},
  author={McCarthy, John},
  year={1964},
  publisher={Stanford University. Computer Science Department. Artificial Intelligence~…}
}

@article{knuth1968semantics,
  title={Semantics of context-free languages},
  author={Knuth, Donald E},
  journal={Mathematical systems theory},
  volume={2},
  number={2},
  pages={127--145},
  year={1968},
  publisher={Springer}
}

@article{avizienis2004basic,
  title={Basic concepts and taxonomy of dependable and secure computing},
  author={Avizienis, Algirdas and Laprie, J-C and Randell, Brian and Landwehr, Carl},
  journal={IEEE transactions on dependable and secure computing},
  volume={1},
  number={1},
  pages={11--33},
  year={2004},
  publisher={IEEE}
}

@article{lyons1962use,
  title={The use of triple-modular redundancy to improve computer reliability},
  author={Lyons, Robert E and Vanderkulk, Wouter},
  journal={IBM journal of research and development},
  volume={6},
  number={2},
  pages={200--209},
  year={1962},
  publisher={IBM}
}

@article{prechelt2002finding,
  title={Finding plagiarisms among a set of programs with JPlag.},
  author={Prechelt, Lutz and Malpohl, Guido and Philippsen, Michael and others},
  journal={J. Univers. Comput. Sci.},
  volume={8},
  number={11},
  pages={1016},
  year={2002}
}

@article{chen2021evaluating,
  title={Evaluating large language models trained on code},
  author={Chen, Mark and Tworek, Jerry and Jun, Heewoo and Yuan, Qiming and Pinto, Henrique Ponde De Oliveira and Kaplan, Jared and Edwards, Harri and Burda, Yuri and Joseph, Nicholas and Brockman, Greg and others},
  journal={arXiv preprint arXiv:2107.03374},
  year={2021}
}

@article{li2022competition,
  title={Competition-level code generation with alphacode},
  author={Li, Yujia and Choi, David and Chung, Junyoung and Kushman, Nate and Schrittwieser, Julian and Leblond, R{\'e}mi and Eccles, Tom and Keeling, James and Gimeno, Felix and Dal Lago, Agustin and others},
  journal={Science},
  volume={378},
  number={6624},
  pages={1092--1097},
  year={2022},
  publisher={American Association for the Advancement of Science}
}

@inproceedings{wang-etal-2025-codecontests,
    title = "{C}ode{C}ontests+: High-Quality Test Case Generation for Competitive Programming",
    author = "Wang, Zihan  and
      Liu, Siyao  and
      Sun, Yang  and
      Ding, Ming  and
      Li, Hongyan",
    editor = "Christodoulopoulos, Christos  and
      Chakraborty, Tanmoy  and
      Rose, Carolyn  and
      Peng, Violet",
    booktitle = "Findings of the Association for Computational Linguistics: EMNLP 2025",
    month = nov,
    year = "2025",
    address = "Suzhou, China",
    publisher = "Association for Computational Linguistics",
    url = "https://aclanthology.org/2025.findings-emnlp.299/",
    doi = "10.18653/v1/2025.findings-emnlp.299",
    pages = "5576--5600",
    ISBN = "979-8-89176-335-7"
}

@article{team2026gemma,
  title={Gemma 4 technical report},
  author={Team, Gemma and Abd, Sherif El and Aggarwal, Vaibhav and Algayres, Robin and Andreev, Alek and Bachem, Olivier and Ballantyne, Ian and Brick, Cormac and C{\u{a}}rbune, Victor and Casbon, Michelle and others},
  journal={arXiv preprint arXiv:2607.02770},
  year={2026}
}

@article{team2024codegemma,
  title={Codegemma: Open code models based on gemma},
  author={Team, CodeGemma and Zhao, Heri and Hui, Jeffrey and Howland, Joshua and Nguyen, Nam and Zuo, Siqi and Hu, Andrea and Choquette-Choo, Christopher A and Shen, Jingyue and Kelley, Joe and others},
  journal={arXiv preprint arXiv:2406.11409},
  year={2024}
}

@article{abdin2024phi,
  title={Phi-4 technical report},
  author={Abdin, Marah and Aneja, Jyoti and Behl, Harkirat and Bubeck, S{\'e}bastien and Eldan, Ronen and Gunasekar, Suriya and Harrison, Michael and Hewett, Russell J and Javaheripi, Mojan and Kauffmann, Piero and others},
  journal={arXiv preprint arXiv:2412.08905},
  year={2024}
}

@article{bai2023qwen,
  title={Qwen technical report},
  author={Bai, Jinze and Bai, Shuai and Chu, Yunfei and Cui, Zeyu and Dang, Kai and Deng, Xiaodong and Fan, Yang and Ge, Wenbin and Han, Yu and Huang, Fei and others},
  journal={arXiv preprint arXiv:2309.16609},
  year={2023}
}

@article{yang2025qwen3,
  title={Qwen3 technical report},
  author={An Yang and Anfeng Li and Baosong Yang and Beichen Zhang and Binyuan Hui and Bo Zheng and Bowen Yu and Chang Gao and Chengen Huang and Chenxu Lv and Chujie Zheng and Dayiheng Liu and Fan Zhou and Fei Huang and Feng Hu and Hao Ge and Haoran Wei and Huan Lin and Jialong Tang and Jian Yang and Jianhong Tu and Jianwei Zhang and Jianxin Yang and Jiaxi Yang and Jing Zhou and Jingren Zhou and Junyang Lin and Kai Dang and Keqin Bao and Kexin Yang and Le Yu and Lianghao Deng and Mei Li and Mingfeng Xue and Mingze Li and Pei Zhang and Peng Wang and Qin Zhu and Rui Men and Ruize Gao and Shixuan Liu and Shuang Luo and Tianhao Li and Tianyi Tang and Wenbiao Yin and Xingzhang Ren and Xinyu Wang and Xinyu Zhang and Xuancheng Ren and Yang Fan and Yang Su and Yichang Zhang and Yinger Zhang and Yu Wan and Yuqiong Liu and Zekun Wang and Zeyu Cui and Zhenru Zhang and Zhipeng Zhou and Zihan Qiu},
  journal={arXiv preprint arXiv:2505.09388},
  year={2025}
}

@misc{qwen25,
      title={Qwen2.5 Technical Report}, 
      author={Qwen and : and An Yang and Baosong Yang and Beichen Zhang and Binyuan Hui and Bo Zheng and Bowen Yu and Chengyuan Li and Dayiheng Liu and Fei Huang and Haoran Wei and Huan Lin and Jian Yang and Jianhong Tu and Jianwei Zhang and Jianxin Yang and Jiaxi Yang and Jingren Zhou and Junyang Lin and Kai Dang and Keming Lu and Keqin Bao and Kexin Yang and Le Yu and Mei Li and Mingfeng Xue and Pei Zhang and Qin Zhu and Rui Men and Runji Lin and Tianhao Li and Tianyi Tang and Tingyu Xia and Xingzhang Ren and Xuancheng Ren and Yang Fan and Yang Su and Yichang Zhang and Yu Wan and Yuqiong Liu and Zeyu Cui and Zhenru Zhang and Zihan Qiu},
      year={2025},
      eprint={2412.15115},
      archivePrefix={arXiv},
      primaryClass={cs.CL},
      url={https://arxiv.org/abs/2412.15115}, 
}

@article{liu2024deepseek,
  title={Deepseek-v2: A strong, economical, and efficient mixture-of-experts language model},
  author={Liu, Aixin and Feng, Bei and Wang, Bin and Wang, Bingxuan and Liu, Bo and Zhao, Chenggang and Dengr, Chengqi and Ruan, Chong and Dai, Damai and Guo, Daya and others},
  journal={arXiv preprint arXiv:2405.04434},
  year={2024}
}

@article{lozhkov2024starcoder,
  title={Starcoder 2 and the stack v2: The next generation},
  author={Lozhkov, Anton and Li, Raymond and Allal, Loubna Ben and Cassano, Federico and Lamy-Poirier, Joel and Tazi, Nouamane and Tang, Ao and Pykhtar, Dmytro and Liu, Jiawei and Wei, Yuxiang and others},
  journal={arXiv preprint arXiv:2402.19173},
  year={2024}
}

@article{jiang2023mistral,
  title={Mistral 7b. arxiv},
  author={Jiang, Albert Q and Sablayrolles, A and Mensch, A and Bamford, C and Chaplot, D Singh and Casas, Ddl and Bressand, F and Lengyel, G and Lample, G and Saulnier, L and others},
  journal={arXiv preprint arXiv:2310.06825},
  volume={10},
  number={3},
  year={2023}
}

@article{grattafiori2024llama,
  title={The llama 3 herd of models},
  author={Grattafiori, Aaron and Dubey, Abhimanyu and Jauhri, Abhinav and Pandey, Abhinav and Kadian, Abhishek and Al-Dahle, Ahmad and Letman, Aiesha and Mathur, Akhil and Schelten, Alan and Vaughan, Alex and others},
  journal={arXiv preprint arXiv:2407.21783},
  year={2024}
}

@article{roziere2023code,
  title={Code llama: Open foundation models for code},
  author={Roziere, Baptiste and Gehring, Jonas and Gloeckle, Fabian and Sootla, Sten and Gat, Itai and Tan, Xiaoqing Ellen and Adi, Yossi and Liu, Jingyu and Sauvestre, Romain and Remez, Tal and others},
  journal={arXiv preprint arXiv:2308.12950},
  year={2023}
}

@inproceedings{utpala-etal-2024-language,
    title = "Language Agnostic Code Embeddings",
    author = "Utpala, Saiteja  and
      Gu, Alex  and
      Chen, Pin-Yu",
    editor = "Duh, Kevin  and
      Gomez, Helena  and
      Bethard, Steven",
    booktitle = "Proceedings of the 2024 Conference of the North American Chapter of the Association for Computational Linguistics: Human Language Technologies (Volume 1: Long Papers)",
    month = jun,
    year = "2024",
    address = "Mexico City, Mexico",
    publisher = "Association for Computational Linguistics",
    url = "https://aclanthology.org/2024.naacl-long.38/",
    doi = "10.18653/v1/2024.naacl-long.38",
    pages = "678--691",
}
